\documentclass{article}
\usepackage[utf8]{inputenc}
\usepackage{main}
\usepackage{microtype}
\usepackage{graphicx}
\usepackage{times}
\usepackage{amsmath}
\usepackage{enumitem}
\usepackage{booktabs}
\usepackage{xcolor}     
\usepackage{natbib}
\definecolor{mydarkblue}{rgb}{0,0.08,0.45}
\usepackage[colorlinks=true,linkcolor=mydarkblue,citecolor=mydarkblue,filecolor=mydarkblue,urlcolor=mydarkblue]{hyperref}
\usepackage{fancyhdr}

\definecolor{gred}{RGB}{250, 210, 207}
\definecolor{coolblue1}{rgb}{0.91, 0.94, 0.98}
\definecolor{coolblue2}{rgb}{0.76, 0.85, 0.94}
\definecolor{coolblue3}{rgb}{0.54, 0.72, 0.87}
\definecolor{coolblue4}{rgb}{1, 1, 1}

\usepackage{xspace}
\usepackage{cleveref}

\usepackage[textsize=tiny]{todonotes}

\usepackage{multirow}
\usepackage{xspace}
\usepackage{wrapfig}
\usepackage{enumitem}
\usepackage{makecell}
\usepackage{caption}
\usepackage{subcaption}
\usepackage[table]{xcolor}

\usepackage[most]{tcolorbox}
\usepackage{etoolbox}
\tcbuselibrary{listingsutf8}

\usepackage{tabularx}
\newcolumntype{L}{>{\raggedright\arraybackslash}X}

\newtcolorbox{LLMPrompt}[1][]{
  enhanced, breakable, verbatim,
  colback=gray!5, colframe=black!15, boxrule=0.4pt, arc=1.5mm,
  left=6pt,right=6pt,top=6pt,bottom=6pt,
  fontupper=\small\ttfamily,
  colbacktitle=gray!50,
  title=#1
}

\iffalse  
    \newcommand{\draftcomment}[3]{}
\else
    \newcommand{\draftcomment}[3]{\textcolor{#2}{{\bf\small [#1: #3]}}}
    
\fi

\newcommand{\correct}{FuncPass\xspace}
\newcommand{\secu}{SecPass\xspace}
\newcommand{\method}{\texttt{AuraForge}\xspace}
\newcommand{\dataset}{\texttt{AuraGym}\xspace}

\newcommand{\tablesize}{\fontsize{8.5}{10}\selectfont}

\definecolor{mediumgreen}{RGB}{0, 150, 50}
\definecolor{mediumred}{RGB}{180, 30, 30}

\usepackage{lineno}

\definecolor{darkblue}{rgb}{0, 0, 0.5}
\hypersetup{colorlinks=true, citecolor=darkblue, linkcolor=darkblue, urlcolor=darkblue}

\begin{document}

\title{\textbf{AuraForge: Scaling Security Supervision for Training Coding Agents}}

\author{
\textbf{Danqing Wang}$^{1}$ \quad
\textbf{Songwen Zhao}$^{1,2}$\thanks{This work was done while
Songwen Zhao and Jierui Wang were interns at Carnegie Mellon
University.} \quad
\textbf{Harsh Sharma}$^{1}$ \quad
\textbf{Jierui Wang}$^{1}$ \\[0.2em]
\textbf{Andre Vicente Duarte}$^{1}$ \quad
\textbf{Ivan Bercovich}$^{3,4}$ \quad
\textbf{Lei Li}$^{1}$ \\[0.6em]
\textsuperscript{1}Carnegie Mellon University \quad
\textsuperscript{2}University of California, Los Angeles \quad
\textsuperscript{3}ScOp Venture Capital. \\
    \href{https://huggingface.co/datasets/dqwang122/AuraGym}{\includegraphics[height=0.4cm]{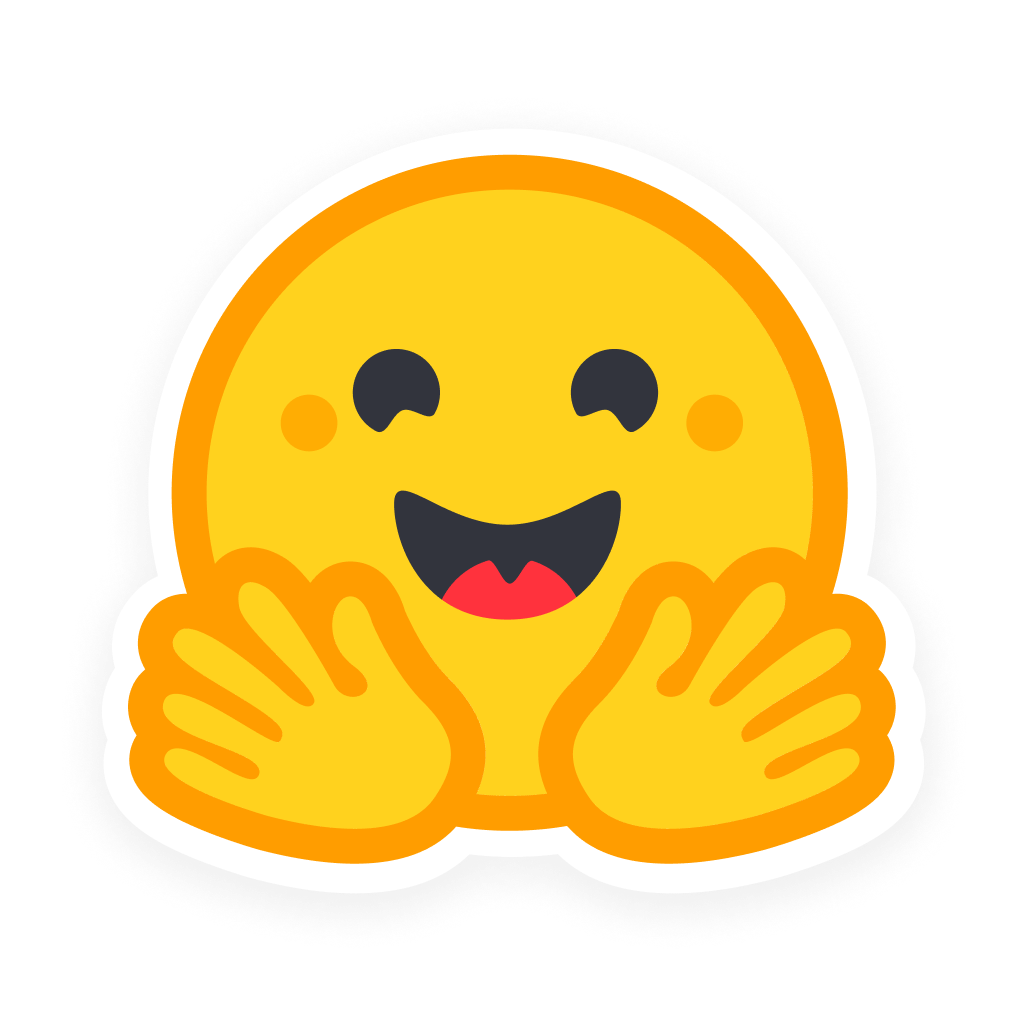} \textbf{dqwang122/AuraGym}}
}

\maketitle
\thispagestyle{fancy}
\fancyhead{}
\lhead{\includegraphics[height=0.5cm]{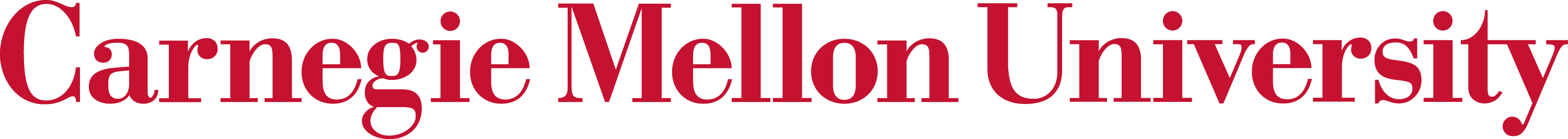}}
\rhead{%
  \raisebox{-0.1cm}{\includegraphics[height=0.8cm]{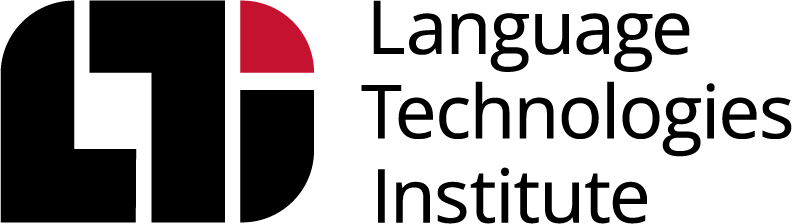}}%
}
\renewcommand{\headrulewidth}{0pt}
\setlength{\headheight}{12pt}
\addtolength{\topmargin}{00pt}
\setlength{\headsep}{3mm}

\vspace{-1.0em}

\begin{abstract}

Coding agents are now proficient enough to generate complex software applications from a single prompt. As their capabilities have grown, human oversight has increasingly shifted from line-by-line code review toward hands-off evaluation of outcomes. However, recent studies have shown that such a transition exposes a critical risk: functional correctness alone does not guarantee a secure implementation. Despite growing attention to code security, training safer coding agents remains challenging because reliable security supervision is difficult to obtain at scale from real-world repositories. 
We introduce \method to synthesize and validate executable security tests for training secure coding agents. 
Our approach combines attack-oriented test synthesis, language-extensible task construction, and safeguards against reward hacking. 
Using \method, we construct \dataset, a multi-language and multi-CWE executable training gym: 679 executable feature-implementation tasks from 344 real-world repositories across Python, JavaScript, and TypeScript, covering 177 CWE categories.
On the subset with human-written security tests, \method produces about 3 times as many test cases on average and reduces the false-positive rate by 83.23\%, allowing alternative secure implementations to receive correct supervision. Training Qwen3.5-4B with synthesized security tests gains larger improvements than human-written security tests (average 19.7 \correct and 6.2 \secu vs. 14.9 \correct and 4.4 \secu) on three languages. 
These results demonstrate that \method provides more diverse and reliable security supervision to train secure coding agents. 
\end{abstract}

\section{Introduction}
\label{sec:intro}

\begin{figure}[ht]
    \centering
    \includegraphics[width=\linewidth]{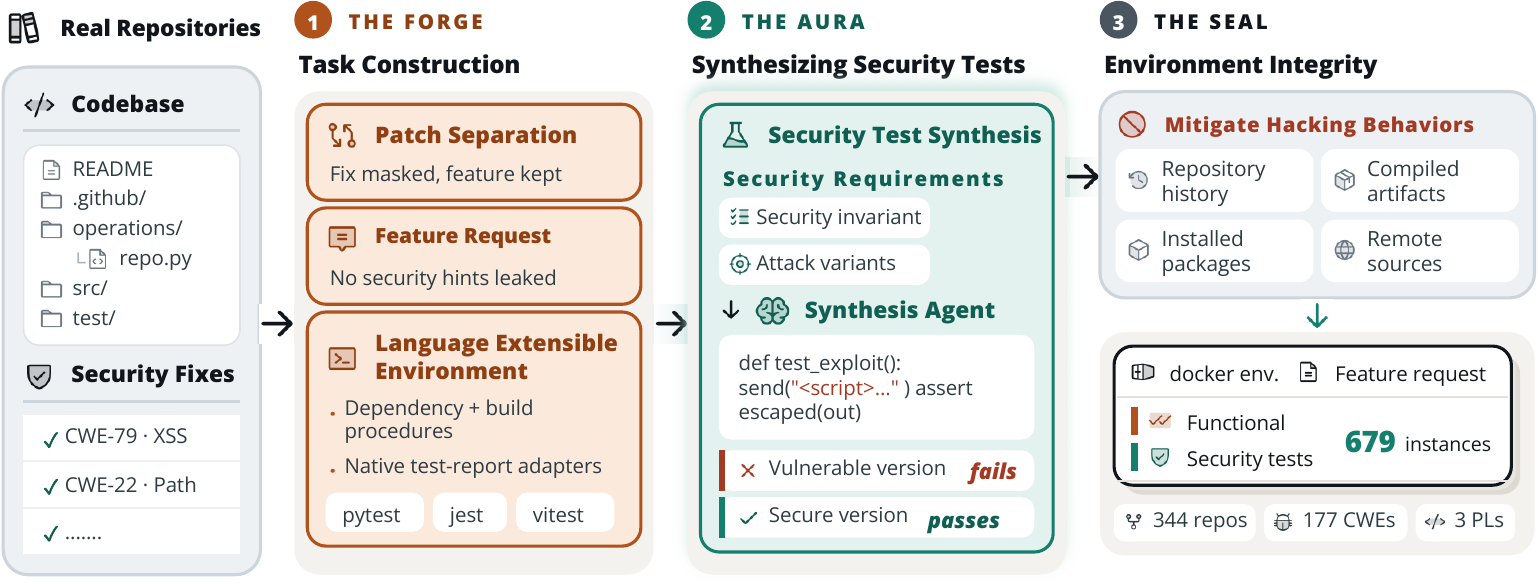}
    \caption{\method. Starting from real repositories and historical security fixes, \emph{The Forge} derives feature requests without leaking security hints, and constructs language-extensible execution environments. \emph{The Aura} synthesizes security tests and validates them. \emph{The Seal} removes implementation-leakage channels. Finally, it produces 679 executable instances from 344 repositories across 177 CWE categories and 3 languages.}
    \label{fig:auraforge}
\end{figure}

Software development is shifting from writing code line by line to describing desired outcomes and letting AI agents realize them: an emerging practice known as \emph{vibe coding}~\citep{fawzy2025vibe,sarkar2025vibe}. Coding agents can create applications, add functionality, fix bugs, refactor components, and modify existing repositories, and they increasingly generate more code than humans can review.
Success is therefore typically judged by whether the software runs and satisfies the user's explicit feature request. However, security requirements are often implicit. A prompt may not state that untrusted input must be sanitized, authorization boundaries preserved, or sensitive data protected, although violating these constraints can turn an otherwise functional implementation into a vulnerability. 
Because human review can no longer be relied upon to catch such violations, security must be taken care of by the agent itself: training vibe coding agents for real-world deployment requires supervision that captures both functional correctness and security.

Recent benchmarks spanning backend application generation, repository-level coding, and feature requests derived from real-world vulnerability-introducing changes consistently reveal a performance gap between functional correctness and security: agents frequently satisfy the requested functionality while producing vulnerable implementations~\citep{vero2025baxbench,chen-etal-2026-securevibebench,zhao2026is}. These findings establish that a stronger general coding capability does not, itself, ensure secure implementation.

Measuring this gap is only a first step, and improving agents requires executable security supervision at training scale. Existing coding-agent training resources largely derive supervision from functional tests~\citep{pantraining,badertdinov2025swerebench,yang2025swesmith}. Historical vulnerability fixes offer a promising source of security-relevant tasks, but scaling them into executable training data introduces 3 bottlenecks. First, reliable security tests are not consistently available: existing human-written security tests for a specific vulnerability fix may miss other vulnerable behavior or reject valid equivalent implementations. Second, task construction must accommodate language-specific source structures, dependency managers, build systems, and test runners. Third, environment construction must ensure that both functional and security tests execute reliably while preventing solution leakage from repository history, generated artifacts, installed package copies, or external sources. Otherwise, an agent may pass the tests by recovering the historical implementation rather than independently producing a secure solution.

At the core of this problem is the quality of the executable security oracle. A useful security test should satisfy three properties: it should (i) \textit{expose the targeted vulnerability when present}, (ii) \textit{accept alternative implementations that satisfy the same security requirement}, and (iii) \textit{execute reproducibly while distinguishing assertion failures from infrastructure errors}. A weak test produces false negatives and provides little useful supervision, and an implementation-specific test produces false positives by encoding incidental details of the historical patch. Test quality must also be considered jointly with task construction: source masking should remove the target logic without breaking the interfaces needed to implement it, and evaluation should exercise the agent's modified source rather than stale or retrieved artifacts.

We study these challenges in repository-level feature implementation with hidden security risks. An agent receives a repository from which the target functionality has been removed, a task description, and an executable environment. The coding agent's implementation is then evaluated on both functional and security tests. 
To address the bottlenecks above, we introduce \method. As shown in \autoref{fig:auraforge}, it synthesizes attack-oriented security tests, constructs tasks through extensible language adapters, and safeguards against reward hacking from retrieval from the internal and external environment.
On the subset where such tests exist, \method produces about $3\times$ as many test cases on average and reduces the false-positive rate by 83.23\%, so that secure implementations differing from the reference are no longer wrongly penalized. Training Qwen3.5-4B~\citep{qwen3.5} with mini-swe-agent~\citep{yang2024sweagent} on synthesized supervision improves functional and security pass@1 by 19.7 and 6.2 percentage points, compared with 14.9 and 4.4 points using human-written tests on three languages. 

Our contributions are as follows:
\begin{itemize}[leftmargin=*,nosep]
    \item We propose \method, which consists of the synthesis of security tests from attacker aspects, adapters to cover different languages, and safeguards for environment integrity.  
    \item Using \method, we construct \dataset with 679 executable coding tasks from 344 repositories across Python, JavaScript, and TypeScript, covering 177 CWE categories.
    \item We show that \method provides more diverse and reliable security supervision than human-written tests, making it more reliable for training secure coding agents. 
\end{itemize}

\section{Related Work}
\label{sec:related}
\textbf{Executable Training Data for Coding Agents.}
Progress in repository-level coding agents has motivated the construction of scalable task collections with reproducible execution environments and test-based rewards. SWE-Gym provides real-world GitHub issues together with executable environments for training agents and verifiers~\citep{pantraining}. SWE-rebench automates the collection of interactive tasks from pull requests and continuously gathers fresh instances to mitigate benchmark contamination~\citep{badertdinov2025swerebench}. Other work synthesizes tasks to scale beyond naturally occurring issues: SWE-Smith constructs training instances by modifying Python repositories to break existing tests~\citep{yang2025swesmith}, while R2E-Gym derives executable tasks from commits using test generation and back-translation and combines execution-based and execution-free verifiers~\citep{jain2025r2egym}. These datasets primarily supervise functional issue resolution. In contrast, our work constructs executable tests that specifically expose security violations and studies their use as training signals for secure coding agents.

\textbf{Secure Coding Benchmark.}
A growing body of work evaluates whether generated code is both functional and secure. BaxBench measures backend application generation using functionality tests and end-to-end exploits~\citep{vero2025baxbench}. 
SecureVibeBench extends this setting to multi-file edits in real repositories, pairing functionality checks with proof-of-concept exploits and static analysis~\citep{chen-etal-2026-securevibebench}. SusVibes focuses on realistic feature requests whose human implementations introduced vulnerabilities, revealing a substantial gap between functional correctness and security in agent-generated patches~\citep{zhao2026is}. AutoBaxBuilder reduces the expert effort required to construct such evaluations by automatically generating application tasks, functionality tests, and security-probing exploits from scratch~\citep{arx2026autobaxbuilder}. While these benchmarks expose security failures, our focus is complementary: we synthesize security tests for real-world repository instances at scale and use them not only for evaluation but also as executable supervision during training. We compare our \dataset with existing works in \autoref{tab:dataset_comparison}.

\begin{table*}[ht]
\centering
\tablesize
\caption{Comparison of general SWE training datasets and secure-coding benchmarks. Instances denote tasks rather than trajectories. A dash indicates that repository counts are not applicable or CWE coverage is not reported.}
\label{tab:dataset_comparison}
\begin{tabular}{llrrrl}
\toprule
\textbf{Dataset} & \textbf{Task setting} & \textbf{Instances} & \textbf{Repos.} & \textbf{CWEs} & \textbf{Languages} \\
\midrule
\multicolumn{6}{c}{\textit{General SWE agent training datasets}} \\
\midrule
SWE-Gym & Issue resolution & 2,438 & 11 & --- & Python \\
SWE-smith & Synthetic bug repair & $\approx$50,000 & 128 & --- & Python \\
SWE-rebench & Issue resolution & 21,336 & 3,468 & --- & Python \\
SWE-rebench V2 & Issue resolution & 32,079 & 3,617 & --- & 20 languages \\
\midrule
\multicolumn{6}{c}{\textit{Secure-coding evaluation benchmarks}} \\
\midrule
BaxBench & Backend generation & 392 & --- & 13 & 6 languages \\
AutoBaxBench & Backend generation & 560 & --- & 11 & 6 languages \\
SecRepoBench & Code completion & 318 & 27 & 15 & C/C++ \\
SecureVibeBench & Repository-level coding & 105 & 41 & 11 & C/C++ \\
SusVibes & Repository-level coding & 186 & 100 & 79 & Python \\
\midrule
\textbf{\dataset{}} & Secure agent training & \textbf{679} & \textbf{344} & \textbf{177} & 3 languages \\
\bottomrule
\end{tabular}
\medskip
\begin{minipage}{\textwidth}
\footnotesize
\end{minipage}
\end{table*}

\textbf{Vulnerability Discovery, Exploitation, and Repair.}
Related benchmarks study other stages of the vulnerability lifecycle. PatchEval evaluates multilingual repair of real-world vulnerabilities and validates a subset of patches with both security and functionality tests~\citep{wei2025patcheval}. CyberGym instead evaluates the generation of proof-of-concept tests that reproduce vulnerabilities in large codebases~\citep{wang2026cybergym}, and CyberGym-E2E expands this scope to the complete workflow of vulnerability discovery, proof-of-concept generation, and patching~\citep{shi2026cybergym}. ExploitGym evaluates whether agents can extend vulnerability-triggering inputs into working exploits across diverse systems~\citep{wang2026exploitgym}, whereas ExploitBench decomposes exploitation into a graded ladder of capabilities rather than treating crashes as successful exploitation~\citep{lee2026exploitbench}. These efforts demonstrate the importance of executable security oracles, but target evaluation of discovery, exploitation, or repair. Our work addresses the data bottleneck by automatically producing diverse security tests that can provide reliable, scalable rewards for training secure coding agents.

\section{\method: Forging Executable Security Tasks at Scale}
\label{sec:method}

\subsection{Problem Formulation}
\label{sec:problem_formulation}

We study repository-level feature implementation with implicit security requirements, following the definition of SusVibes~\citep{zhao2026is}. The request describes the desired functionality without disclosing the historical vulnerability or prescribing its fix. The agent must therefore satisfy both explicit functional requirements and implicit security requirements. 

We define a training instance as $x = (\mathcal{R}, d, \mathcal{T}_{\mathrm{func}}, \mathcal{T}_{\mathrm{sec}}, \mathcal{E})$, where $\mathcal{R}$ is the repository state, $d$ is the feature request, $\mathcal{T}_{\mathrm{func}}$ and $\mathcal{T}_{\mathrm{sec}}$ are functional and security test suites, and $\mathcal{E}$ specifies the executable environment. Separately, construction metadata $\mathcal{Z}$ records information such as historical vulnerable $\mathcal{S}^*_{\mathrm{vuln}}$, fixed implementations $\mathcal{S}^*_{\mathrm{sec}}$ and their vulnerability provenance. These artifacts support the task and the test construction and are withheld from the coding agent during the evaluation. 

Functional tests assess the requested behavior, while security tests assess targeted security properties. Good security tests should satisfy three requirements: (i) \emph{detection}, exposing the targeted vulnerability when present; (ii) \emph{implementation independence}, accepting alternative implementations that satisfy the same security requirement; and (iii) \emph{execution reliability}, producing reproducible outcomes while distinguishing test failures from infrastructure errors. Passing the tests provides evidence of correctness within their scope, rather than a guarantee of security.

\textbf{Challenges in scaling executable security supervision.}
Scaling this real-world repository-level setting into a diverse training resource presents three challenges. First, \textbf{\textit{reliable security tests}} that satisfy the three requirements above are not consistently available. Historical fixes may lack security tests, while existing tests may miss vulnerable behavior or reject valid alternative implementations.
Second, \textbf{\textit{multilingual task construction}} must accommodate differences in code structure, dependency management, build procedures, and test reporting and parsing. 
Third, \textbf{\textit{environment integrity}} requires executable environments that preserve legitimate development capabilities while preventing unintended access to reference implementations through repository history, compiled artifacts, installed packages, and remote sources. Without these controls, passing tests may reflect reference retrieval rather than secure feature implementation.

To address these challenges, we propose \method, a scalable pipeline that synthesizes reliable security tests and constructs executable tasks across programming languages, providing supervision for training more secure vibe coding.

\subsection{Synthesizing Reliable Security Tests}
\label{sec:security_supervision}

Given $(\mathcal{R}, d, \mathcal{T}_{\mathrm{func}}, \mathcal{E})$ and the construction metadata $\mathcal{Z}$, \method synthesizes a security test suite $\mathcal{T}_{\mathrm{sec}}$. A solution is judged secure if it fails no test in $\mathcal{T}_{\mathrm{sec}}$. To achieve detection without sacrificing implementation independence, \method synthesizes tests from the attacker's perspective, targeting the effects of attacks rather than the location or mechanism of the historical fix. It then validates the tests in $\mathcal{E}$ for execution reliability and for their ability to distinguish $\mathcal{S}^*_{\mathrm{vuln}}$ from $\mathcal{S}^*_{\mathrm{sec}}$.

\subsubsection{Attack-Oriented Synthesis}
Synthesis proceeds in two phases. The identification phase represents the security requirement as a \emph{security invariant}, an effect that must always hold, and a set of \emph{attack variants} that cover distinct entry points, input encodings, and code paths. Each variant specifies an attack surface, an observable effect, and a sourced payload when available. A risk narrative further describes the attacker's capabilities and objectives.

The synthesis phase then exercises each attack variant through the feature's attack surface and asserts that its effect is prevented. A feature may span multiple components, from the frontend to the backend database. Unit tests tied to a security guard in one component can reject alternative implementations that enforce equivalent protection elsewhere, producing false positives. Asserting on attack effects instead admits different defense locations and mechanisms: unauthorized access may be prevented by either redirection or rejection, and an injection defense need not produce an exact escaped string as long as no executable \texttt{<script>} element is rendered. Assertions also exclude identified security-irrelevant differences, such as output formatting. More details in Appendix~\ref{app:synthesis_details}.

\subsubsection{Execution-based validation.}
To ensure execution reliability, we instruct the synthesis agent to avoid external network dependencies and order-dependent execution, to seed and bound any fuzzing, and to handle vulnerability-induced hangs through explicit assertions with bounded timeouts. Import, dependency, setup, connection, and infrastructure-timeout errors are never treated as security evidence, and the agent abstains when it cannot construct a suite that meets these requirements.

To ensure detection, the suite must distinguish $\mathcal{S}^*_{\mathrm{vuln}}$ from $\mathcal{S}^*_{\mathrm{sec}}$ in $\mathcal{E}$, and this outcome serves as execution feedback during synthesis. We then independently execute each test on both implementations in fresh environments outside the synthesizer's control. A suite is accepted only if both runs complete without test errors and at least one test fails on $\mathcal{S}^*_{\mathrm{vuln}}$ and passes on $\mathcal{S}^*_{\mathrm{sec}}$.

\subsection{Language-Extensible Construction}
\label{sec:dataset_construction}

Each task masks a feature implementation from the repository and asks the agent to restore it from the feature request. Constructing such tasks involves three steps whose details depend on the language: transforming the source, preparing an environment in which tests exercise that source, and interpreting test outcomes. \method splits each step into a \emph{shared core}, which is language-agnostic and reused unchanged, and a \emph{language adapter}, which supplies only the language-dependent parts. Supporting a new language therefore requires implementing the three adapter components, without modifying the core workflow or its validation logic. We describe each component below, using our JavaScript/TypeScript adapter as a running example.

\begin{figure}[th]
    \centering
    \includegraphics[width=\linewidth]{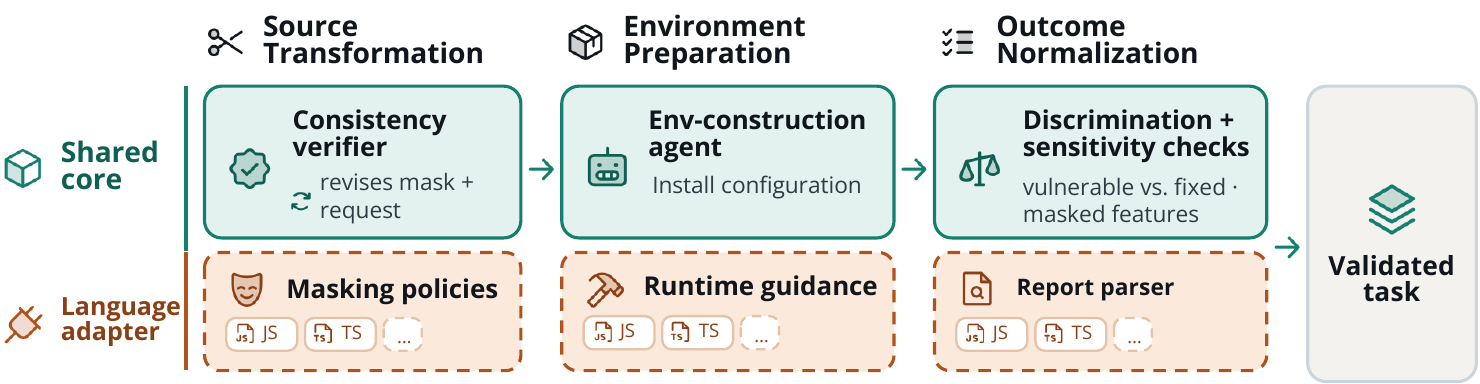}
    \caption{Division between the shared core and the language adapter in \method.}
    \label{fig:adapter}
\end{figure}

\textbf{Source transformation.}
Feature masking must remove implementation logic without unintentionally changing the interfaces needed to complete the task. Because these boundaries depend on language syntax, \method uses language-specific policies for patch classification and masking, coupled with a shared consistency verifier. In JavaScript and TypeScript, the policies account for implementation bodies, initializers, exports, and type annotations. Verification feedback guides revisions to the mask and feature request.

\textbf{Environment preparation.}
Repository tooling can decouple the source an agent edits from the artifacts tests actually execute, so a passing run does not by itself establish valid supervision. A shared environment-construction agent inspects repository configuration and documentation to determine installation, build, and test procedures, guided by language-specific runtime and build instructions from the adapter. For JavaScript and TypeScript, these instructions cover runtime and package-manager selection, workspace dependencies, and regeneration of compiled outputs, so that tests exercise the current source rather than stale builds.

\textbf{Outcome normalization.}
Test runners report outcomes in different formats, which complicates comparison across repository states. The adapter parses runner-specific reports into a common result format that separates passes, assertion failures, and execution errors. The shared core then applies the same checks to every language: security-test discrimination between the vulnerable and fixed states, and functional-test sensitivity to the masked state. For JavaScript and TypeScript, the parser handles nested workspace runners and multiple summary blocks.

\subsection{Preserving Environment Integrity}
\label{sec:environment_integrity}

Feature masking alone does not guarantee that an agent must implement the requested functionality: equivalent implementations may remain accessible through local artifacts or external sources. Reward hacking is widespread among agentic benchmarks~\citep{bercovich2026terminal,stein2026detecting}, including mining git history for historical fixes on SWE-bench-style benchmarks, and it persists even when agents are explicitly instructed not to cheat~\citep{metr-2025-recent-reward-hacking}. Such access undermines the training signal by allowing retrieval to substitute for implementation.
To avoid that, we remove these sources proactively so that retrieval is not possible in our executable environment.

\begin{wrapfigure}{r}{0.46\textwidth}
\vspace{-10pt}
\centering
\includegraphics[width=0.46\textwidth]{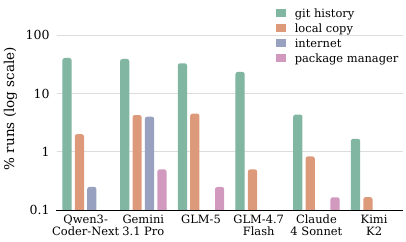}
\caption{Retrieval rates of SusVibes runs, per model and channel (runs pooled over scaffolds), with the repository history present and the network reachable.}
\label{fig:hack_rate}
\vspace{-10pt}
\end{wrapfigure}

\textbf{Measuring implementation retrieval.}
To quantify this risk in our setting, we evaluated 26 agent--model combinations on SusVibes in environments that retained the repository history and allowed network access, while the prompt explicitly prohibited retrieving the reference implementation. We flagged potential retrieval with a high-recall rule-based matcher, and an LLM judge then examined each flagged trajectory to determine whether the agent located and used the reference implementation. Across 5{,}174 runs, retrieval occurred in 16\% of runs and in up to 40\% of one model's runs (Figure~\ref{fig:hack_rate}), predominantly through git history lookups, consistent with prior findings~\citep{endorlabs2026asl,stein2026detecting}. Beyond these known channels, agents also recovered the upstream package from local copies in \texttt{site-packages}, including within their own scaffold's Python environment.

\textbf{Controlling implementation access.}
Selectively blocking leaking commits is infeasible: a masked feature typically evolves across many historical commits by different developers, and in preliminary experiments agents continued to exploit the remaining history even after the commits containing the reference implementation were hidden. We therefore remove the git history entirely and ship each task repository as a single initial commit. We also remove local copies of the target implementation from the solving environment, including build artifacts, dependency caches, and installed packages. 
Appendix~\ref{app:environment_integrity} gives the audit protocol and the sanitization rules. During solving, internet and package-manager access is blocked at the command level (Appendix~\ref{app:trajectory_distillation}).
\method's solving and evaluation run in separate environments. Security tests are introduced only at evaluation time in a fresh container, and any edits to files used by the tests are discarded before the solution is applied, so that a solution cannot tamper with the tests that judge it.

\section{\dataset: A Multi-Language, Multi-CWE Security Gym}
\label{sec:dataset}

We collect vulnerability-fixing commits from MoreFixes~\citep{10.1145/3663533.3664036} and the GitHub Advisory Database (GHSA)~\citep{ghsa_repo}, covering Python and extending to JavaScript and TypeScript. 
Unlike SusVibes~\citep{zhao2026is}, which is limited to Python commits accompanied by human-written security tests, \method synthesizes security tests and supports new languages through language adapters (Appendix~\ref{app:js_ts_collection}). The resulting dataset, \dataset, comprises 679 instances from 344 repositories spanning 177 CWE categories. We also hold out an additional validation set for the new TypeScript/JavaScript SusVibes$_{tsjs}$. This
split is repository-disjoint and emphasizes vulnerabilities fixed in 2026 to reduce memorization and contamination (Appendix~\ref{app:js_ts_split}).

\textbf{Comparison with existing agentic coding datasets.}
\autoref{tab:dataset_comparison} compares \dataset with general SWE training datasets and secure coding benchmarks. General SWE training datasets provide large-scale functional supervision but lack security requirements. 
Secure coding benchmarks evaluate security across tasks such as backend generation, code completion, and repository-level implementation, but their limited size and CWE coverage make them insufficient for training. 
Among these benchmarks, \dataset contains the most instances and covers over $3\times$ as many repositories and $2\times$ as many CWE categories as SusVibes, the most diverse prior benchmark. The next largest by instance count, AutoBaxBench, contains 560 instances drawn from 40 backend scenarios across 14 frameworks, but covers only 11 CWE categories. It is constructed with AutoBaxBuilder~\citep{arx2026autobaxbuilder}, which uses LLMs to generate backend scenarios, functional tests, and security exploits, with manual verification of each scenario. However, it is restricted to synthetic REST API backends and the predefined CWE classes of BaxBench, covering 11 CWE categories. In contrast, \method derives tasks from real-world vulnerability fixes, capturing the complexity of existing codebases and enabling broader repository and CWE diversity.

\begin{figure*}[ht]
\centering
\begin{minipage}[t]{0.32\textwidth}
    \centering
    \includegraphics[width=\linewidth]{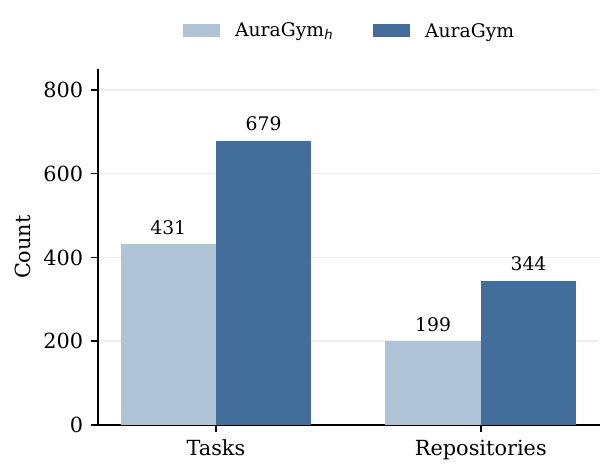}
    \caption*{(a) Dataset scale}
\end{minipage}
\hfill
\begin{minipage}[t]{0.32\textwidth}
    \centering
    \includegraphics[width=\linewidth]{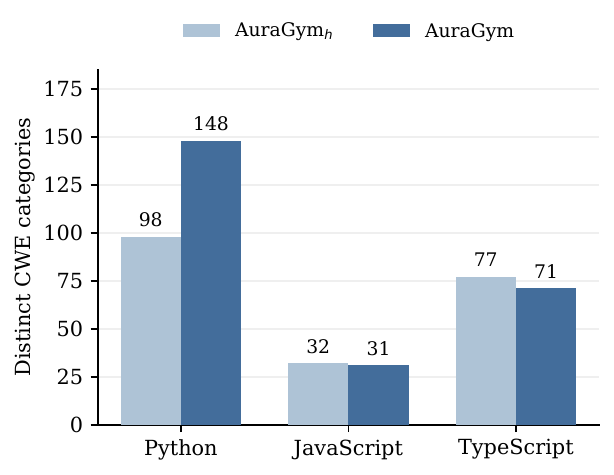}
    \caption*{(b) CWE diversity}
\end{minipage}
\hfill
\begin{minipage}[t]{0.32\textwidth}
    \centering
    \includegraphics[width=\linewidth]{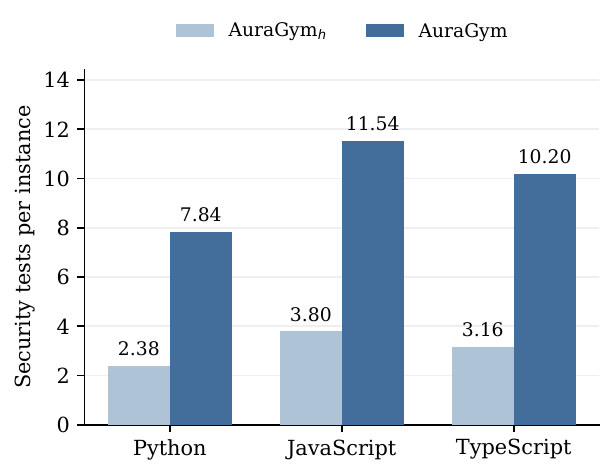}
    \caption*{(c) Security tests}
\end{minipage}
\caption{Comparison of human-written \dataset$_h$ and synthesized \dataset{}: (a) task and repository counts; (b) distinct CWE categories by language; and (c) average security tests per task. }
\label{fig:dataset-statistics}
\end{figure*}

\textbf{Comparison with human-written tests.}
To compare synthesized and human-written supervision, we construct $\dataset_{h}$, the subset of instances whose source vulnerability-fix commits include human-written security tests. \autoref{fig:dataset-statistics} compares \dataset with $\dataset_{h}$ in scale, CWE diversity, and the number of security tests, with full statistics in \autoref{tab:benchmark_statistics}. 
Because it depends on the availability of human-written security tests, $\dataset_{h}$ is substantially smaller and covers fewer CWE categories. Moreover, \dataset provides on average $3\times$ as many security tests per instance as $\dataset_{h}$, yielding denser security supervision. This increase holds across Python, JavaScript, and TypeScript, indicating that the synthesis pipeline generalizes across languages and testing frameworks.

\section{Experiments and Analysis}
\label{sec:analysis}
Our experiments address two research questions: \textit{RQ1:  How does the quality of \method's synthesized security tests compare with that of human-written tests? RQ2: Does \dataset provide effective training signals for secure vibe coding?}

To answer these questions, we collect agent trajectories on \dataset using mini-swe-agent~\citep{yang2024sweagent} with two models, Muse Spark 1.3~\citep{metaimuse2026spark13} and DeepSeek-V4.1-Flash~\citep{deepseekai2026deepseekv41flash}, under two prompting settings: a generic security reminder and task-specific security hints. This provides 3{,}753 valid trajectories from Muse Spark 1.3 and 2{,}218 from DeepSeek-V4.1-Flash. For RQ1, we use the grading outcomes of these trajectories to assess how well the synthesized security tests distinguish secure from vulnerable implementations. For RQ2, we fine-tune Qwen3.5-4B~\citep{qwen3.5} on trajectories that are both functionally correct and secure, and evaluate whether this improves the model's functional and security performance. Details of trajectory collection are provided in Appendix~\ref{app:trajectory_distillation}.

\subsection{RQ1: Synthesized Tests Match Human Recall, 83\% Fewer False Positives}
\label{sec:gen_vs_human}

We evaluate the reliability of synthesized security tests by comparing them with human-written tests on two metrics: \emph{vulnerability detection recall}, the fraction of vulnerable solutions a suite rejects, and \emph{false positive rate}, the fraction of secure solutions a suite rejects. 

We grade trajectories on the 347 tasks in $\dataset_h$ with both the human-written and the \method-synthesized security suites, and retain the 2{,}450 solutions that pass the functional tests. We then compare the two suites' security verdicts on these solutions. Table~\ref{tab:gen_vs_human} reports the two metrics by language, and Figure~\ref{fig:wrong_verdicts} counts every wrong verdict by the kind of test error behind it. Appendix~\ref{app:gen_vs_human} details the procedure and provides the categorization of incorrect verdicts (Table~\ref{tab:gen_vs_human_categories}), an audit of what each suite asserts, a breakdown by vulnerability family, and five case studies.

\begin{table}[t]
    \begin{minipage}[t]{0.555\textwidth}
        \vspace{0pt}
        \centering
        \caption{Synthesized vs.\ human-written security tests, graded on the same functionally correct agent solutions. Recall is the share of vulnerable solutions a suite rejects; FPR is the share of secure solutions it rejects.}
        \label{tab:gen_vs_human}
        \footnotesize
        \setlength{\tabcolsep}{2.5pt}
        \begin{tabular*}{\linewidth}{@{\extracolsep{\fill}}lrrrcccc@{}}
            \toprule
            \multirow{2}{*}{Language} & \multirow{2}{*}{Tasks} & \multicolumn{2}{c}{Solutions} & \multicolumn{2}{c}{Recall (\%) $\uparrow$} & \multicolumn{2}{c}{FPR (\%) $\downarrow$} \\
            \cmidrule(lr){3-4}\cmidrule(lr){5-6}\cmidrule(lr){7-8}
            & & sec. & vuln. & Hum. & Synth. & Hum. & Synth. \\
            \midrule
            Python     & 155 & 366 &   547 & 96.2 & 98.0 & 12.8 & \textbf{2.2} \\
            TypeScript & 151 & 190 & 1,059 & 99.2 & 98.9 & 21.1 & \textbf{3.7} \\
            JavaScript &  45 &  79 &   254 & 100 & 99.2 & 24.1 & \textbf{3.8} \\
            \midrule
            \textbf{Overall} & 347 & 635 & 1,815 & 98.3 & 98.6 & 16.7 & \textbf{2.8} \\
            \bottomrule
        \end{tabular*}
    \end{minipage}\hfill
    \begin{minipage}[t]{0.42\textwidth}
        \vspace{0pt}
        \centering
        \includegraphics[width=\linewidth]{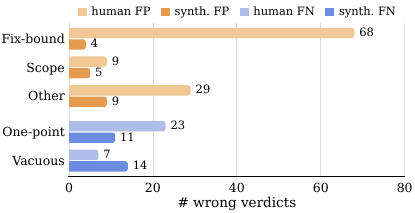}\par
        \captionsetup{skip=2pt}
        \captionof{figure}{Kinds of every wrong verdict behind Table~\ref{tab:gen_vs_human}. Fix-bound tests demand a detail of the fix; one-point tests try one input or path; vacuous tests never observe the vulnerable behaviour (full in Table~\ref{tab:gen_vs_human_categories}).}
        \label{fig:wrong_verdicts}
    \end{minipage}
\end{table}

Both suites return the same verdict on 2{,}271 of the 2{,}450 solutions (93\%), and we adopt this shared verdict as the ground-truth security label. For the remaining 179 solutions, we determine the ground truth by executing an attack against each solution in its environment, with the rulings verified by hand (Appendix~\ref{app:gvh_labels}). Table~\ref{tab:gen_vs_human} reports each suite's vulnerability detection recall and false positive rate against these labels, broken down by language.

\textbf{\method's synthesized security tests achieve vulnerability detection recall comparable to human-written tests.}
Human-written tests set a demanding bar for detection, as they were written by maintainers with knowledge of the vulnerability and its exploit. Across 1{,}815 vulnerable solutions, the synthesized suites reject 98.6\%, and the human-written suites 98.3\%, and the two remain within two percentage points of each other in every language (Table~\ref{tab:gen_vs_human}). However, they miss vulnerabilities for different reasons (Figure~\ref{fig:wrong_verdicts}, bottom). Most of the human-written suites' misses arise from tests that exercise only a single input or code path, allowing a solution that closes only that hole to pass (one-point: 23 of 30). For the synthesized suites, this share is 11 of 25, all due to single inputs; none stem from an untested code path, reflecting the enumeration of attack variants across entry points, encodings, and code paths (Section~\ref{sec:security_supervision}). Their remaining misses arise from checks that never observe the solution's vulnerable behavior, mostly due to mocking (vacuous: 14 of 25).

\textbf{Synthesized security tests reduce the false positive rate by 83\%.}
High recall alone does not make a useful oracle: a suite that rejects every solution achieves perfect recall but provides no learning signal. We therefore measure how often each suite rejects solutions adjudicated as secure. The human-written suites reject 16.7\% of the 635 secure solutions, compared with 2.8\% for the synthesized suites, and the gap holds in all languages (13--24\% vs.\ 2--4\%). Figure~\ref{fig:wrong_verdicts} (top) shows where it lies: almost entirely in tests that demand a detail of the historical fix (fix-bound: 68 human vs.\ 4 synthesized), while tests that require protection the vulnerability does not call for (scope) and the remaining causes (other) are few on both sides. On 33 of the 183 tasks with at least one secure solution (18\%), the human-written suites reject \emph{every} secure solution; the synthesized suites do so on only 4 (2\%).

\textbf{Human-written tests fail secure solutions because they assert the fix, not the absence of the attack.}
Of the human-written suites' 106 false positives, 68 (64\%, across 32 tasks) come from tests that would fail any correct fix other than the historical one, because they require the fix's exact output (32\%), its particular safe reaction (19\%), its error message (7\%), or a helper function defined only in the fix (7\%). For example, a Markdown converter is vulnerable to resource exhaustion when a heading tag with an extreme level, such as \texttt{<h9999999>}, causes it to generate millions of characters. The historical fix clamps the level to six, and its test asserts that \texttt{<h10>} is rendered as \texttt{<h6>}. A solution that instead renders such tags as plain text prevents the attack equally well but fails the test.

The synthesized suites produce only 4 false positives of this kind. An audit of the security assertions confirms this difference: 87\% of synthesized suites assert that the attack's effect is absent, compared with 48\% of human-written suites, and human-written suites that assert the fix's outcome reject secure solutions 1.7$\times$ as often as those that assert the attack's effect (20.7\% vs.\ 12.0\%; Appendix~\ref{app:assertion_audit}). By vulnerability family, human-written false positive rates are highest for prototype pollution (33\%) and resource exhaustion (19\%), where multiple defense mechanisms are secure but the tests assert the historical one (Appendix~\ref{app:gvh_family}).

\begin{table*}[ht]
    \centering
    \caption{Functional and security performance on the SusVibes (Python) and newly created TypeScript/JavaScript evaluation sets. SusVibes is the original Python version from ~\citet{zhao2026is} and SusVibes$_{tsjs}$ is the test split for the new TS and JS.  Nemotron 3.5 is the  Lightning version.}
    \label{tab:sft_results}
    \small
    \begin{tabular}{lccccc}
        \toprule
        \multirow{2}{*}{Model}
            & \multirow{2}{*}{Size}
            & \multicolumn{2}{c}{SusVibes (186)}
            & \multicolumn{2}{c}{SusVibes$_{tsjs}$ (82)} \\
        \cmidrule(lr){3-4}\cmidrule(lr){5-6}
            & & \correct & \secu & \correct & \secu \\
        \midrule
        GPT 5.6 Sol             & --       & 84.40 & 21.50 & 92.70 & 18.30 \\
        Muse Spark 1.3          & --       & 79.57 & 15.59 & 82.90 & 19.50 \\
        GLM 4.7 Flash & 30B-A3B  & 24.73 & 4.30 & 24.39 & 4.88 \\
        Nemotron 3.5   & 30B-A3B  & 15.05 & 3.23 & 25.61 & 3.66 \\
        \midrule
        Qwen 3.5 4B & 4B & 12.90 & 2.10 & 12.19 & 0.00 \\
        \quad + $\dataset_{h}$ &  4B & 22.04 & 4.84 & 32.93 & 6.10 \\
        \quad + \dataset         &  4B & \textbf{27.96} & \textbf{5.91} & \textbf{36.59} & \textbf{8.54} \\
        \bottomrule
    \end{tabular}
\end{table*}

\subsection{RQ2: Synthesized Signals Outperform Human-Written Tests for Training}
\label{sec:analysis_scale}

We use the trajectories of DeepSeek's functionally correct and secure solutions for supervised training because they contain more reasoning content than Muse Spark. 
We use Qwen 3.5 4B as the base model and compare the base checkpoint with two fine-tuned variants: one trained on trajectories selected using the synthesized tests in \dataset and one trained on trajectories selected by $\dataset_{h}$.

\textbf{Evaluation Setup.}
We evaluate all models on SusVibes and TypeScript/JavaScript splits containing 186 and 82 instances using mini-swe-agent. We report \correct, the percentage of instances that pass functional tests, and \secu, the percentage that passes both the functional and security tests. 
During evaluation, we also prohibit commands like \textit{git clone or pip download} to avoid hacking. 

\textbf{Training on \dataset improves both functionality and security, outperforming the 30B-parameter GLM 4.7 Flash and Nemotron 3.5 and narrowing the gap to frontier models.}
Table~\ref{tab:sft_results} shows that training Qwen3.5-4B on \dataset improves \correct and \secu on SusVibes, more than doubling the base model on both metrics. The gains are larger on TS/JS, where \correct increases by 24.40 points and \secu rises from 0.00 to 8.54. 
It also surpasses GLM 4.7 Flash and Nemotron 3.5 on every metric in both benchmarks. 
To separate security from functional gains, we consider the fraction of functionally correct solutions that are also secure (\secu/\correct). Qwen3.5-4B + \dataset reaches 21.1\% on SusVibes and 23.3\% on TS/JS, on par with Muse Spark 1.3 (19.6\% and 23.5\%) and GPT 5.6 Sol (25.5\% and 19.7\%). 

\textbf{Synthesized supervision outperforms human-written supervision.}
Training on $\dataset_h$ with human-written security tests also improves performance, but by a smaller margin. Averaged over both benchmarks, \dataset improves \correct and \secu by 19.7 and 6.2 points, compared with 14.9 and 4.4 points for $\dataset_h$. Combined with the lower false positive rate of synthesized tests (\S\ref{sec:gen_vs_human}), these results indicate that \method provides both more scalable and more reliable security supervision for training.

\section{Conclusion}
\label{sec:conclusion}
As coding agents write more code than humans can review, security must be learned by the agent rather than enforced through human inspection. We present \method, a pipeline that turns real-world vulnerability fixes into executable training tasks with attack-oriented security tests, language-extensible task construction, and safeguards against reward hacking. The resulting \dataset spans 679 tasks, three languages, and 177 CWE categories. Our synthesized tests match the detection recall of human-written tests while reducing false positives by 83\%, and training on \dataset improves both functional and security performance beyond human-written supervision. Future work includes using these executable tests as rewards for reinforcement learning and extending coverage to more languages and vulnerability classes.

\newpage
\subsection*{AI use statement}

We used LLMs to refine the writing and assist with figures and tables. We did not use LLMs to formulate mathematical claims, develop or write proofs, or translate research materials. Other tasks requiring disclosure were not applicable to this work.
The authors reviewed all AI-assisted content and take full responsibility for the final paper, including all text, claims, analyses, and artifacts produced with generative AI assistance.




\subsection*{Reproducibility statement}


We describe the task formulation, security-test synthesis, multilingual task construction, environment safeguards, and trajectory collection in Section \ref{sec:method}. The appendix provides additional details on data sources, filtering criteria, language-specific adaptations, trajectory collection, and fine-tuning. Our experiments specify the functional and security evaluations used to assess supervision quality and coding-agent performance. 



\section*{Acknowledgments}

Danqing Wang acknowledges support from the Amazon AI PhD Fellowship. This work was supported by Microsoft through the Carnegie Mellon CyLab partnership program. We acknowledge financial support from Coefficient Giving and Ivan Bercovich of ScOp Venture Capital.

\bibliography{iclr2027_conference}
\bibliographystyle{iclr2027_conference}

\newpage
\appendix
\label{sec:appendix}

\section{Language Adaptation and Task-Construction Details}
\label{app:language_adaptation}

This section describes the components adapted to language and repository requirements within the shared construction workflow. JavaScript and TypeScript serve as concrete examples of source-transformation, environment-preparation, and test-execution adaptations.

\subsection{Patch Separation and Feature Masking}

Given a historical vulnerability-fixing commit, the pipeline separates implementation changes from test-related changes. File paths provide an initial classification, supplemented by LLM-based classification for ambiguous edits. In particular, dependency and configuration changes may affect application behavior, test setup, or both, requiring inspection beyond their filenames.

Let $\mathcal{C}_{-1}$ denote the vulnerable repository state associated with the fix. A mask $\mathcal{M}$ removes the target feature implementation, producing the task repository $\mathcal{R}=\mathcal{C}_{-1}^{\mathcal{M}}$, together with a corresponding feature request $d$. Language-specific masking policies specify which implementation elements may be removed and which surrounding interfaces should remain. For JavaScript and TypeScript, these policies target implementation bodies and initializers while accounting for declarations, exports, and type annotations.

A shared verifier checks whether the reference implementation satisfies the generated request and whether the mask introduces unrelated changes. Security protections relevant to the feature remain within scope even when the request does not explicitly mention them. Rejection explanations guide revisions to the mask and request.

\subsection{Runtime, Dependency, and Build Configuration}

The environment-construction agent uses repository configuration files, CI workflows, and installation documentation to determine the required runtime, dependencies, and execution commands. For JavaScript and TypeScript, this includes selecting compatible Node.js versions, configuring the appropriate package manager, and resolving dependencies among workspace packages.

Build procedures must preserve consistency between the source state being validated and the artifacts executed by tests. Where tests consume generated outputs, stale artifacts are removed and the relevant project is rebuilt before validation. This prevents an earlier compiled implementation from determining the outcome of a test run against modified source code.

\subsection{Test Execution and Result Normalization}

Repository-specific adapters execute native test workflows and translate their outputs into a common result format. The same adapter is used across validation states to make their outcomes comparable. This is particularly relevant to JavaScript and TypeScript repositories that combine workspace orchestration with individual test runners, producing nested reports or multiple summaries.

Result interpretation distinguishes completed test outcomes from failures of the execution procedure. An error message printed by a test does not by itself establish that the test failed, and an incomplete run cannot be treated as a valid security judgment. Adapters must therefore interpret outcomes according to the repository's testing workflow rather than relying solely on isolated log strings.

\subsection{Execution-Based Task Validation}

The construction workflow compares fixed and vulnerable implementations with and without the test patch, together with the feature-masked vulnerable implementation under the existing functional tests. These comparisons assess whether added security tests distinguish vulnerable and fixed implementations and whether functional tests detect removal of the target feature.

Environment executability and task validity are checked separately. A completed run may contain meaningful assertion failures, whereas an execution error leaves the corresponding comparison unresolved. Applying the same validation criteria across languages provides a common basis for accepting tasks constructed through different source-transformation policies and repository tooling.

\subsection{JavaScript and TypeScript Collection and Filtering}
\label{app:js_ts_collection}

We describe the collection and static filtering of JavaScript and TypeScript candidates. The resulting pool feeds subsequent task construction and execution-based validation.

\paragraph{Data sources.}
We collect candidates from the June 20, 2026 MoreFixes database backup, containing 50,986 records across 10,377 repositories, and GHSA advisories~\citep{ghsa_repo} in a frozen OSV npm snapshot, yielding 3,742 unique retrieved commits. These source counts represent database records and unique commits, respectively.

\paragraph{MoreFixes filtering.}
We export records with a commit--CVE relevance score of at least 65, a CVE identifier year of 2014 or later, and nonempty changes to existing non-test \texttt{.js}, \texttt{.ts}, or \texttt{.tsx} files with unchanged paths. Deduplicating CVE--commit--repository tuples yields 3,189 records. Static screening removes 1,628 records without qualifying tests, 455 without qualifying implementation code, 132 with implementation patches exceeding 500 lines or 10 files, and 7 containing only test or configuration changes. Further deduplication and date, runtime, and primary-language filtering retain 249 candidates (Table~\ref{tab:morefixes_filtering}).

\begin{table}[t]
\centering
\small
\caption{MoreFixes filtering. Removed denotes the reduction from the preceding row; dashes indicate unreported counts.}
\label{tab:morefixes_filtering}
\begin{tabular}{p{0.49\linewidth}rrr}
\toprule
Selection stage & Candidates & Removed & Repos. \\
\midrule
Initial JS/TS export & 3,189 & --- & --- \\
Implementation, test, and patch-size filters & 967 & 2,222 & --- \\
Candidate deduplication & 909 & 58 & 514 \\
Fixing-commit date $\geq$ 2024-01-01 & 311 & 598 & --- \\
Identified Node.js major $\geq$ 18 & 255 & 56 & 127 \\
Repository primary language is JS/TS & 249 & 6 & --- \\
\bottomrule
\end{tabular}
\end{table}

\paragraph{GHSA filtering.}
We extract direct GitHub commit references, excluding withdrawn advisories and malware records. We retain single-parent commits with reconstructable diffs and complete required metadata, including CWE information, and exclude commits containing file additions, deletions, or renames. Requiring JS/TS changes and test files, followed by date, runtime, and primary-language filtering, leaves 817 candidates. Removing 203 overlaps with MoreFixes yields 614 candidates (Table~\ref{tab:ghsa_filtering}).

\begin{table}[t]
\centering
\small
\caption{GHSA filtering and cross-source deduplication. Removed denotes the reduction from the preceding row; dashes indicate unreported counts.}
\label{tab:ghsa_filtering}
\begin{tabular}{p{0.49\linewidth}rrr}
\toprule
Selection stage & Candidates & Removed & Repos. \\
\midrule
Retrieved unique commits & 3,742 & --- & --- \\
Patch structure, parent, and metadata filters & 2,371 & 1,371 & --- \\
Patch includes JS/TS & 2,204 & 167 & --- \\
Patch includes test files & 1,389 & 815 & 489 \\
Fixing-commit date $\geq$ 2024-01-01 & 896 & 493 & --- \\
Identified Node.js major $\geq$ 18 & 817 & 79 & 198 \\
Repository primary language is JS/TS & 817 & 0 & 198 \\
Cross-source deduplication & 614 & 203 & --- \\
\bottomrule
\end{tabular}
\end{table}

\paragraph{Shared selection policies.}
Both sources require fixing commits dated on or after January 1, 2024 and an identified Node.js major version of at least 18, inferred from the target commit's CI configuration, version files, Docker configuration, \texttt{engines} declarations, and documentation. Runtime filtering excludes 28 below-threshold and 28 unknown-version MoreFixes candidates, and 40 below-threshold and 39 unknown-version GHSA candidates. Cross-source deduplication uses normalized repository identity and fixing-commit SHA, assigning overlaps to MoreFixes.
The final pool contains 863 candidates: 249 from MoreFixes and 614 from GHSA. These candidates are used as the input of the task construction, leads to 229 instances in total.

\subsection{TypeScript/JavaScript Training and Test Split}
\label{app:js_ts_split}

We construct a repository-disjoint split using the vulnerability-fix year from \texttt{cve\_fix\_date} and the repository identity from \texttt{project}. Candidate test repositories are non-OpenClaw repositories containing at least one record from 2026. Once selected, an entire repository is assigned to the test set, including any of its records from 2024 or 2025; all 112 records from \texttt{openclaw/openclaw} remain in training.

We select test repositories through binary optimization. The primary objective minimizes the number of test records while requiring coverage of all 72 labels in the candidate pool: 71 specific CWE identifiers and the \texttt{NVD-CWE-noinfo} placeholder. Among equally small solutions, a secondary objective minimizes the number of pre-2026 test records. The proven minimum contains 82 test instances, of which 69 (84.1\%) are from 2026; a 72-instance test set cannot satisfy repository disjointness and complete label coverage simultaneously. Multi-label instances may cover more than one CWE, so label coverage does not imply one test instance per label.

Every instance outside the selected test repositories is assigned to training. The resulting split is therefore repository-disjoint and emphasizes recent vulnerabilities to reduce memorization and contamination, but is not strictly chronological because the training set also contains records from 2026.

\section{Security-Test Synthesis Details}
\label{app:synthesis_details}

This section gives the two stages of Section~\ref{sec:security_supervision} in enough detail to reproduce their design decisions.

\paragraph{Stage 1: the security property.}
An agent researches the vulnerability from the public record only (the CVE entry, advisories, and what they link to), with web search and no access to the repository, and returns a structured security property. Its fields are chosen so that nothing in it names an implementation: the concrete weakness; a risk narrative in plain language (who the attacker is, what they can do, what they gain); the invariant, stated as an effect that must always hold; the attack surface, the entry points the attacker controls; the attack variants, each a distinct way to reach the effect through one of those entry points, with what is observable from outside when it succeeds and, where the record supplies one, a verbatim payload with its source; and the security-irrelevant differences, details that vary between correct implementations and must not be asserted. The agent also writes, for each statement, why it judges the attack rather than one way of stopping it, and lists what the public record left open. Defences are deliberately not enumerated: public sources describe few of them, and naming one would steer the tests toward it.

\paragraph{Stage 2: the tests.}
A second agent runs inside the task's execution environment with the repository in its vulnerable state and writes the test suite. It receives the security property, the feature request (as the functional specification, not as a list of behaviours to test), and the vulnerable feature's code. It is instructed to cover every attack variant, to assert the observable effect of each variant rather than any mechanism that might prevent it, and to assert nothing tied to the security-irrelevant differences. The historical fix is available as a patch file that the agent can apply and revert; its role is an execution signal, letting the agent check that its tests fail on the vulnerable state and pass on the fixed one, not a source of assertions. The functional test command is given only so that the tests run under the same interpreter and dependencies.

\paragraph{Reliability rules.}
Tests must avoid network access and order-dependent execution, seed and bound any fuzzing, and turn a hang caused by the vulnerability into an assertion failure under a timeout, so that a denial-of-service test concludes rather than stalls. Import, dependency, setup, connection and infrastructure-timeout errors never count as security evidence. An agent that cannot write a suite meeting these rules submits nothing, and the task is dropped rather than given a weak oracle.

\paragraph{Post-processing and acceptance.}
From the agent's diff we keep only added test files: binary files and any change to a file the task's implementation patch touches are removed, so a suite cannot alter the code it grades. The suite is then run, outside the agent's control and in fresh environments, on the historical vulnerable and fixed implementations; it is accepted only if both runs complete without test errors and at least one test fails on the vulnerable implementation and passes on the fixed one. This check uses a single pair of implementations, so it is necessary rather than sufficient; Section~\ref{sec:gen_vs_human} measures how the accepted suites behave on many other implementations.

\section{Environment Integrity: Retrieval Audit and Sanitization}
\label{app:environment_integrity}

This appendix supports Section~\ref{sec:environment_integrity}: how retrieval of the reference implementation was detected in the SusVibes audit, why a copy of the fixed code survives inside a task image, and how sanitization was verified. Table~\ref{tab:retrieval_channels} lists the four retrieval channels and the countermeasure that closes each.

\begin{table}[h]
\centering
\small
\begin{tabular}{@{}l >{\raggedright\arraybackslash}p{6.3cm} >{\raggedright\arraybackslash}p{4.2cm}@{}}
\toprule
Channel & How the reference is obtained & Countermeasure \\
\midrule
git history & \texttt{git show}, \texttt{git log -p}, or \texttt{git checkout} on a pre-mask revision & history stripped; the task ships as a single commit \\
local copy & reading a copy already on disk: \texttt{site-packages}, a build directory, a vendored tree & image sanitized \\
internet & fetching the upstream file or archive from a code host & command filtering while solving \\
package manager & installing the upstream package at its fixed version, then reading its source & command filtering while solving \\
\bottomrule
\end{tabular}
\caption{Retrieval channels observed in the audit and how each is closed.}
\label{tab:retrieval_channels}
\end{table}

\paragraph{Detecting retrieval.}
A run counts as retrieval when the agent views the reference implementation of the masked feature, or content directly derived from it, by any route; reading the current tree, installing the declared dependencies, and reading the functional tests do not count. Detection has two stages because no single rule is both complete and precise. A rule-based scan of every tool call flags any read of a non-HEAD revision, any fetch from a code host, any pinned package install, and any read under an installed-package or vendored path; it favours recall and flags about a quarter of the runs. A language-model verifier then reads each flagged trajectory in an isolated directory with read-only tools and rules whether the reference was actually obtained and through which channel, citing the steps. The channel is the route by which the content first entered the trajectory, so copying an already fetched file is not a second channel, and a metadata-only \texttt{git log} is not flagged, since hashes and subjects reveal nothing until a later command reads them.

\paragraph{Why the fixed code survives in the image.}
A task image is built with the repository at its fixed state, so that dependencies resolve and build steps succeed, and only the git work tree is then rolled back and masked. Everything the build produced from the fixed tree stays: \texttt{build/lib}, the installed copy in \texttt{site-packages} or a tox virtualenv, a wheel in the pip cache, compiled extensions, and for JavaScript the \texttt{dist} output. Sanitization removes a copy when it is this project's own build output or installed copy, and leaves it when it belongs to a dependency: a library that vendors the same code is part of the environment, not of our build. pip itself is the extreme case, since it vendors \texttt{requests}, so a fix to requests ships inside every pip; there the only remedy is removing pip, which is acceptable when the test command does not use it.

\paragraph{Verification.}
Each image is swept by regenerating the fixed content of every masked file and comparing it against every file in the image, which finds copies nobody named. After removal the task is re-validated: the security tests must still fail on the vulnerable state and pass on the fixed state, the functional baseline must match the published record, and the test command must be unchanged; a removal that breaks any of these is not applied. An artifact the tests cannot run without therefore stays, and a task whose image still exposes its fix is excluded. In every leaky image we examined, the test command already executed the agent's work tree rather than the leaked copy: the defect was an answer key readable on disk, not a harness scoring the wrong code, so sanitization changes no existing score.

\section{More Details on Existing Agentic Coding Datasets}
\label{app:comparison_datasets}
We summarize the datasets in Table~\ref{tab:dataset_comparison}, distinguishing general SWE training resources from secure-coding evaluation benchmarks. Instance counts refer to tasks rather than collected interaction trajectories. Repository counts describe source-project diversity, while CWE counts describe reported vulnerability categories; differences in task construction and CWE annotation limit direct comparisons of these quantities.

\paragraph{General SWE training datasets.}
SWE-Gym provides 2,438 executable issue-resolution tasks from 11 Python repositories for training and evaluating coding agents. SWE-smith scales task construction through synthetic bugs, producing approximately 50,000 tasks across 128 Python repositories. SWE-rebench collects 21,336 issue-resolution tasks from 3,468 Python repositories. SWE-rebench V2 extends this approach to multilingual construction, with a main containerized collection of 32,079 tasks from 3,617 repositories across 20 programming languages. These resources support learning repository-level software engineering through executable functional feedback, without explicitly organizing supervision around vulnerability categories.

\paragraph{Backend-generation benchmarks.}
BaxBench and AutoBaxBench evaluate whether agents can generate functionally correct and secure backends from specifications. BaxBench contains 28 scenarios instantiated across 14 frameworks, yielding 392 tasks across six programming languages and 13 CWE categories. AutoBaxBench, constructed using AutoBaxBuilder, contains 40 scenarios instantiated across the same number of frameworks, yielding 560 tasks across six languages. It reports 11 CWE categories but excludes CWE-400 from its main evaluation. These instance counts include framework-specific versions of scenarios; repository counts are not applicable because the tasks concern backend generation from specifications.

\paragraph{Repository-based security benchmarks.}
SecRepoBench evaluates secure code completion with 318 tasks from 27 C/C++ repositories covering 15 CWE categories. SecureVibeBench evaluates repository-level secure implementation with 105 tasks from 41 C/C++ projects covering 11 CWE categories. SusVibes evaluates feature implementation with implicit security requirements, comprising 186 tasks from 100 Python repositories spanning 79 CWE categories. SusVibes is particularly relevant to our task formulation: agents must satisfy functional requirements while avoiding vulnerabilities that are not explicitly identified in the feature request.

\begin{table*}[ht]
    \centering
    \caption{Statistics of the synthesized $\dataset_{s}$ and the human-written comparison set $\dataset_{h}$. The numbers of functional tests, security tests, and target files are reported as per-instance averages. An instance whose repository mixes JavaScript and TypeScript is counted under both languages.}
    \label{tab:benchmark_statistics}
    \footnotesize
    \begin{tabular*}{\textwidth}{@{\extracolsep{\fill}}llcccccc}
        \toprule
        Version & Language & \# Instances & \# Repos & \# CWEs & \makecell{Avg. \#\\target files} & \makecell{Avg. \#\\func. tests} & \makecell{Avg. \#\\security tests} \\
        \midrule
        \multirow{4}{*}{$\dataset$}
            & Python     & 476 & 266 & 148 & 1.65 & 73.22 &  7.84 \\
            & JavaScript &  46 &  36 &  31 & 1.70 & 32.22 & 11.54 \\
            & TypeScript & 161 &  46 &  71 & 1.99 & 20.52 & 10.20 \\
        \cmidrule(lr){2-8}
            & \textbf{Overall}    & 679 & 344 & 177 & 1.72 & 53.54 & 8.62 \\
        \midrule
        \multirow{4}{*}{$\dataset_{h}$}
            & Python     & 202 & 109 &  98 & 1.39 & 89.51 & 2.38 \\
            & JavaScript &  53 &  42 &  32 & 1.79 & 32.22 & 3.80 \\
            & TypeScript & 182 &  54 &  77 & 2.04 & 20.52 & 3.16 \\
        \cmidrule(lr){2-8}
            & \textbf{Overall}    & 431 & 199 & 139 & 1.69 & 48.56 & 2.90 \\
        \bottomrule
    \end{tabular*}
\end{table*}

\begin{table*}[ht]
\centering
\caption{Statistics of the trajectories collected by DeepSeek-V4.1-Flash and Muse Spark 1.3 that are functionally correct and secure. Coverage denotes the percentage of instances represented by at least one
  retained trajectory within each model's instance pool (679 for DeepSeek-V4.1-Flash; 710 for Muse Spark 1.3, which was collected on an earlier version of \dataset). Token counts are reported in units of 1,024
  tokens.}
\label{tab:sft_trajectories}
\small
\setlength{\tabcolsep}{3pt}
\begin{tabular*}{\textwidth}{@{\extracolsep{\fill}}lllccccc@{}}
\toprule
Language & Model & Setting & \# Traj. & \# Inst. & Coverage & Avg. turns & Avg. tokens (K) \\
\midrule
\multirow{6}{*}{\textbf{Python}}
& \multirow{3}{*}{DeepSeek}
& generic & 392 & 163 & 34.24\% & 39.75 & 52.15 \\
& & security-hint & 77 & 77 & 16.18\% & 46.53 & 59.84 \\
& & \textbf{overall} & \textbf{469} & \textbf{226} & \textbf{47.48\%} & \textbf{40.86} & \textbf{53.42} \\
\cmidrule(l){2-8}
& \multirow{3}{*}{Muse Spark}
& generic & 462 & 152 & 31.67\% & 42.91 & 35.59 \\
& & security-hint & 161 & 112 & 23.33\% & 47.22 & 43.27 \\
& & \textbf{overall} & \textbf{623} & \textbf{264} & \textbf{55.00\%} & \textbf{44.02} & \textbf{37.57} \\
\midrule
\multirow{6}{*}{\textbf{TypeScript}}
& \multirow{3}{*}{DeepSeek}
& generic & 61 & 33 & 21.02\% & 52.00 & 69.18 \\
& & security-hint & 26 & 26 & 16.56\% & 63.35 & 79.75 \\
& & \textbf{overall} & \textbf{87} & \textbf{54} & \textbf{34.39\%} & \textbf{55.39} & \textbf{72.34} \\
\cmidrule(l){2-8}
& \multirow{3}{*}{Muse Spark}
& generic & 86 & 36 & 20.34\% & 68.74 & 62.39 \\
& & security-hint & 13 & 12 & 6.78\% & 57.92 & 41.22 \\
& & \textbf{overall} & \textbf{99} & \textbf{48} & \textbf{27.12\%} & \textbf{67.32} & \textbf{59.61} \\
\midrule
\multirow{6}{*}{\textbf{JavaScript}}
& \multirow{3}{*}{DeepSeek}
& generic & 33 & 15 & 35.71\% & 33.42 & 42.03 \\
& & security-hint & 5 & 5 & 11.90\% & 27.40 & 36.45 \\
& & \textbf{overall} & \textbf{38} & \textbf{20} & \textbf{47.62\%} & \textbf{32.63} & \textbf{41.30} \\
\cmidrule(l){2-8}
& \multirow{3}{*}{Muse Spark}
& generic & 37 & 13 & 27.66\% & 50.16 & 37.82 \\
& & security-hint & 4 & 4 & 8.51\% & 37.75 & 30.81 \\
& & \textbf{overall} & \textbf{41} & \textbf{17} & \textbf{36.17\%} & \textbf{48.95} & \textbf{37.13} \\
\midrule
\multirow{6}{*}{\textbf{Total}}
& \multirow{3}{*}{DeepSeek}
& generic & 486 & 211 & 31.08\% & 40.86 & 53.60 \\
& & security-hint & 108 & 108 & 15.91\% & 49.69 & 63.55 \\
& & \textbf{overall} & \textbf{594} & \textbf{300} & \textbf{44.18\%} & \textbf{42.46} & \textbf{55.41} \\
\cmidrule(l){2-8}
& \multirow{3}{*}{Muse Spark}
& generic & 585 & 201 & 28.31\% & 47.17 & 39.67 \\
& & security-hint & 178 & 128 & 18.03\% & 47.79 & 42.84 \\
& & \textbf{overall} & \textbf{763} & \textbf{329} & \textbf{46.34\%} & \textbf{47.31} & \textbf{40.41} \\
\bottomrule
\end{tabular*}
\end{table*}

\section{\dataset Trajectory Collection}
\label{app:trajectory_distillation}
We collect secure feature-implementation trajectories from Muse Spark 1.3~\citep{metaimuse2026spark13} and DeepSeek v4.1 Flash~\citep{deepseekai2026deepseekv41flash}.
Using the mini-swe-agent~\citep{yang2024sweagent}, we first collect agent--environment interaction trajectories with the original feature request $d$. To increase the coverage of tasks with successful trajectories, we additionally augment requests with security hints that guide the coding agent toward secure implementations. These hints include the original CWE or CVE description from the metadata of each instance. 
Before training, we remove the added hints from the task prompts, restoring $d$ as the input.

During collection, the coding agent interacts with the sanitized solving environment described in Section~\ref{sec:environment_integrity}. To mitigate reward hacking through external implementation retrieval, we apply command filtering to block online search and code-retrieval operations, including \texttt{git clone}, \texttt{git remote}, and package retrieval through \texttt{npm}, \texttt{yarn}, and \texttt{pip download}. We adopt mini-swe-agent since it requires no scaffold packages in the task-solving container, which would otherwise be a cheating source. We retain only trajectories whose resulting implementations pass both $\mathcal{T}_{\mathrm{func}}$ and $\mathcal{T}_{\mathrm{sec}}$. 

Finally, we get 3,753 valid trajectories from Muse and 2,218 trajectories from DeepSeek v4.1 Flash, excluding empty solution patches and hacked behaviors. 
For supervised fine-tuning, we only keep trajectories filtered by functional and security tests. The statistics are listed in \autoref{tab:sft_trajectories}.
During fine-tuning, we limit the context to 64K, use a batch size of 8 with 3 epochs. The peak learning rate is 1e-5. 

\section{Comparing Synthesized and Human-Written Security Tests}
\label{app:gen_vs_human}

This appendix gives the full method behind Section~\ref{sec:gen_vs_human}: how the solutions were collected, how the labels were obtained and verified, the finer error categories with examples, the split by vulnerability family, and the limits of the comparison.

\subsection{Tasks and Solutions}
\label{app:gvh_tasks}

The comparison uses every task of $\dataset_h$ that carries both suites: the human-written security tests (the test files the maintainers committed together with the vulnerability fix) and the synthesized tests of \method\ for the same task. There are 434 such tasks, 165 in Python and 269 in TypeScript/JavaScript. Both suites graded the solutions that two coding agents (Muse Spark 1.3 and a DeepSeek model) produced on these tasks while collecting trajectories: 3,190 trajectories in total. A solution enters the comparison when both suites ran to completion and left logs, and when it passes the repository's functional tests under both gradings; the same patch on the same task counts once. This leaves 2,533 solutions on 351 tasks. The largest exclusion is flakiness of the repository's own functional suite (351 trajectories where one of the two gradings failed it), not a difference between the security suites.

\paragraph{Recorded verdicts that were not test outcomes.}
Before trusting a recorded pass, we scanned every log for the runner's own summary line. Thirty-six Python solutions carried a recorded \emph{pass} from a suite while a security test of that suite had in fact failed in the run: 32 on the human side, where the grader's parser missed the runner's summary line in a less common format (for example \texttt{FAILED (failures=1, errors=1)}) or its count-based rule tolerated the failure, and 4 on the synthesized side, where the count-based rule tolerated failures within the published baseline. One TypeScript solution carried a human pass the log could not confirm. These 37 were dropped, so that neither suite is charged with a verdict the harness rather than the tests produced.

\subsection{Labels}
\label{app:gvh_labels}

\paragraph{Agreement.}
The two suites agree on 2,271 of the 2,450 solutions kept (511 both pass, 1,760 both fail). We take the shared verdict as the label. This is the one place the comparison is not grounded by execution: a vulnerable solution both suites accept, or a secure one both reject, is invisible to it, so recall is measured against what at least one suite catches.

\paragraph{Adjudication of disagreements.}
We adjudicate each disagreeing solution (up to four per task; 35 above that cap were not adjudicated) with an LLM judge (Claude Opus, run through the Claude Agent SDK) that works inside the solution's own evaluation container with the solution applied and is told nothing about which suite is which. The judge (i) researches the vulnerability from the CVE record and public advisories, (ii) writes its own attack against the solution's outermost entry point and shows it succeeds on the historical vulnerable code and fails on the historical fix, (iii) runs it against the solution, (iv) runs both test suites, shown only as ``suite A'' and ``suite B'' with no recorded verdict, and (v) rules whether the solution is secure and which suite is right, with a rationale, a confidence, and the mechanism by which the wrong suite failed. We discard a ruling when the judge's own run of a suite contradicts the recorded verdict (11 solutions, all traced to a grading artifact rather than to the suite). This produced 190 rulings on 88 tasks.

\paragraph{Human verification.}
We verified the rulings by hand: we read every medium-confidence ruling (13) and 20 of the high-confidence ones (there were none at low confidence), re-deriving the verdict from the judge's attack transcript and both suites' outputs. Thirty-one held. We overturned two and recorded the reason: a Django open-redirect solution the judge had ruled vulnerable for rejecting fewer leading characters than upstream, where the characters it accepts all resolve to same-origin paths (secure); and a nodemailer solution the judge had ruled secure after payloads that all placed the closing quote directly against the \texttt{@}, where a quoted local part followed by whitespace is misrouted exactly as on the vulnerable code (vulnerable). All rates in Section~\ref{sec:gen_vs_human} use the verified labels.

\paragraph{Rates.}
For each suite, the false positive rate is the share of solutions labelled secure that it fails and recall is the share labelled vulnerable that it fails; intervals are 95\% percentile bootstraps over tasks (2,000 resamples). Of the 2,450 solutions, 635 are labelled secure and 1,815 vulnerable. Pooled, the human suites' false positive rate is 16.7\% [12.0, 22.0] and the synthesized suites' 2.8\% [1.0, 5.0]; recall is 98.3\% [97.2, 99.2] and 98.6\% [97.7, 99.4]. By language, the false positive rates are 12.8 [7.3, 18.7] against 2.2 [0.2, 4.8] (Python), 21.1 [11.8, 31.8] against 3.7 [0.6, 7.6] (TypeScript) and 24.1 [7.5, 42.1] against 3.8 [0.0, 13.2] (JavaScript); recall is 96.2 [93.1, 98.6] against 98.0 [95.7, 99.8], 99.2 [98.1, 99.8] against 98.9 [97.8, 99.7], and 100 against 99.2 [97.9, 100]. A task in both TypeScript and JavaScript counts under each language, as in Table~\ref{tab:benchmark_statistics}. The TypeScript/JavaScript sets are dominated by solutions both suites fail (1,245 of 1,568), so their recall is high on both sides and the comparison there rests mainly on the false positive rate; the JavaScript secure side is 79 solutions on 45 tasks, hence its wide intervals. The 1,730 solutions from one agent and the 720 from the other give the same reading (human FPR 17.5\% and 15.5\%, synthesized 1.8\% and 4.4\%).

\subsection{Why Each Suite Was Wrong}
\label{app:gvh_why}

We read every one of the 179 wrong verdicts (106 human false positives, 30 human false negatives, 18 synthesized false positives, 25 synthesized false negatives) against the judge's rationale and our verification notes, and assigned each to one of the categories of Table~\ref{tab:gen_vs_human_categories}, fixed in advance, each a way the security test's own check can be wrong; a verdict that fits none of them (a functional demand in the test file, a flaky or drifted run, a timing or memory threshold) or more than one at once is ``other''. A verdict whose failing tests fall in two categories of the same kind takes the category most of them fall in (4 rows, all fix-bound: a throw and an exported helper, an exact list and a warning text). Table~\ref{tab:gen_vs_human_categories} lists the categories with one example each, grouped into four kinds, each named by the question the test fails, plus ``other''. Of the 29 human false positives under ``other'', 22 are a functional demand in the test file or a flaky run, and 7 have failing tests of two kinds: 4 a fix-bound demand with a functional one, and one each a threshold with a stricter rule, a threshold with a fix-bound demand, and a functional demand with a stricter rule.

\begin{table}[p]
    \centering
    \caption{Causes of wrong verdicts. Every wrong verdict behind Table~\ref{tab:gen_vs_human} is assigned, from the adjudication record, to the category of test error that produced it; the categories are grouped into kinds, each named by the question the test fails. Counts are wrong verdicts (human / synthesized): secure solutions rejected for the false-positive kinds, vulnerable solutions accepted for the false-negative kinds; kind rows give the total, with tasks in parentheses. Each example is a real case; those in Section~\ref{app:case_studies} are referenced.}
    \label{tab:gen_vs_human_categories}
    \footnotesize
    \setlength{\tabcolsep}{3pt}
    \begin{tabular*}{\textwidth}{@{\extracolsep{\fill}}p{0.21\textwidth}p{0.17\textwidth}p{0.43\textwidth}rr@{}}
        \toprule
        Kind: the question the test fails & Category: the test\ldots & Example & Human & Synth. \\
        \midrule
        \multicolumn{5}{@{}l}{\textit{Rejects a secure solution (false positive): the test demands more than that the attack has no effect}} \\
        \midrule
        \multirow{5}{=}{\textbf{Fix-bound}: does the test demand a detail of the historical fix that security leaves free?}
          & \ldots wants the exact form of a correct result & A converter must render an over-long heading tag as \texttt{<h6>}, as the fix does; a solution that renders it as plain text stops the attack just as well (Appendix~\ref{app:case_markdownify}); likewise a query must have the exact SQL text the fix sends (Appendix~\ref{app:case_nocobase}). & 34 & 1 \\
          & \ldots wants one particular safe reaction where another exists & A model loader must refuse a file in safe mode; a solution instead loads it through a restricted loader that cannot run code (Appendix~\ref{app:case_keras}). & 20 & 2 \\
          & \ldots wants the error's wording or type & The error must say ``arbitrary code execution''; a solution refuses the same file with a different message (Appendix~\ref{app:case_keras}). & 7 & 1 \\
          & \ldots calls a helper only the fix defines & The test imports a validator function the fix added; a solution that validates inline has no function of that name. & 7 & 0 \\
        \cmidrule(l){2-5}
          & \textit{all fix-bound} & & \textbf{68} (32) & 4 (3) \\
        \midrule
        \textbf{Scope}: does the test require protection the vulnerability does not call for? & \ldots requires protection against a related weakness that is not this vulnerability & The vulnerability lets a logged-out user act on any account; the test also refuses a logged-in user acting on another user's account. & 9 (4) & 5 (3) \\
        \midrule
        \textbf{Other} & \ldots fails for a reason not above & A functional demand in the test file (the only failing test wants a new input type accepted), a flaky or machine-dependent run (a memory threshold the fix itself exceeds on some machines), or failing tests of more than one category in the same run. & 29 (14) & 9 (3) \\
        \midrule
        \multicolumn{5}{@{}l}{\textit{Accepts a vulnerable solution (false negative): the test never shows the attack failing on this solution}} \\
        \midrule
        \multirow{3}{=}{\textbf{One-point}: is the solution secure where the test looks and vulnerable next to it?}
          & \ldots tries one input, which the solution blocks while a neighbouring input still works & The test attacks a link-account step while logged out; a solution refuses that attacker and still lets any other logged-in user through (Appendix~\ref{app:case_sentry}); likewise a merge is attacked through one argument position only (Appendix~\ref{app:case_defu}). & 13 & 11 \\
          & \ldots exercises one code path, while the bug sits in another & The test sends a token in its compact form; the solution bounds that path and leaves the JSON form unbounded. & 10 & 0 \\
        \cmidrule(l){2-5}
          & \textit{all one-point} & & \textbf{23} (13) & 11 (8) \\
        \midrule
        \multirow{3}{=}{\textbf{Vacuous}: can the test's check observe the vulnerable behaviour at all?}
          & \ldots replaces the vulnerable step with a fake, so that step never runs & The test replaces the HTTP client with a fake that never follows a redirect; the bug is in following redirects. & 2 & 6 \\
          & \ldots has a check that cannot fail on this solution & The test counts any exception as ``attack blocked''; the solution lets the traversal through and raises an unrelated error afterwards. & 5 & 8 \\
        \cmidrule(l){2-5}
          & \textit{all vacuous} & & 7 (3) & \textbf{14} (6) \\
        \bottomrule
    \end{tabular*}
\end{table}

The four kinds are not equally deep. Fix-bound tests, one-point tests, and the vacuous tests that rest on a mock (12 of 21) are one root cause seen three ways: the test was written against one implementation. Anchored on the fix's diff, it fails other correct fixes; anchored on the advisory's input, it passes a solution that closes only that hole; a mock shaped to the reference's code path never fires on a solution that takes another. The human suites carry this in the assertion and the input; the synthesized suites carry it in the mock. Nine of the synthesized suites' 14 vacuous checks are on one repository whose code talks to external services, so this failure kind is concentrated rather than spread over the set. Scope is a separate problem: the human suites have 9 such false positives and the synthesized suites 5, so this kind does not depend on who wrote the test but on how far a task's security requirement extends, for example whether a sibling bug the CVE does not name is in scope. So are the timing and memory thresholds grouped under ``other'' (9 of the synthesized suites' 18 false positives, on three denial-of-service tasks): they fail the historical fix itself on some runs, and a different oracle, a refusal or a bound on the work done rather than a threshold, removes them outright.

\subsection{Assertion Audit}
\label{app:assertion_audit}

The categories above are read from the wrong verdicts. To check the mechanism behind the false-positive gap on the suites themselves, we audited every security suite in the comparison on one property, independently of how it graded. The population is the 183 tasks with at least one secure solution, that is, every task where a false positive is possible: 366 suites, of which two human suites contain no security test and are left out. Readers were given the feature request, the historical fix and both test diffs, not the grades, and classified each suite's security tests under a fixed rubric: whether the assertions check that the attacker's effect is absent (a canary not created, a secret not returned, output bounded, an action refused in any form) or check the outcome the historical fix produces (a specific exception class or message, an exact output, an emitted call, a helper's name). Three cases with known answers from the case studies were read first and came back as expected. Table~\ref{tab:gen_vs_human_audit} gives the result.

\begin{table}[t]
    \centering
    \caption{Assertion audit of both suites on the 183 tasks with a secure solution. Each suite's security tests were classified, blind to the grades, by what they assert: that the attack's effect is absent, or the outcome the historical fix produces; mixed suites are counted by their majority. FPR is over the secure solutions of the tasks in that class.}
    \label{tab:gen_vs_human_audit}
    \small
    \begin{tabular*}{\textwidth}{@{\extracolsep{\fill}}lcccc@{}}
        \toprule
        \multirow{2}{*}{Suites that\ldots} & \multicolumn{2}{c}{Human} & \multicolumn{2}{c}{Synthesized} \\
        \cmidrule(lr){2-3}\cmidrule(lr){4-5}
        & share of tasks & FPR & share of tasks & FPR \\
        \midrule
        assert the attack's effect is absent & 48\% & 12.0\% & \textbf{87\%} & 3.0\% \\
        assert the outcome the fix produces  & 52\% & \textbf{20.7\%} & 13\% & 2.0\% \\
        \bottomrule
    \end{tabular*}
\end{table}

Synthesized suites assert the effect in 87\% of tasks, human suites in 48\%, in every language; within the human suites, outcome-asserting ones fail secure solutions 1.7 times as often as effect-asserting ones. Holding the class fixed, the gap between the suites remains (on the 82 tasks where both suites assert the effect, the human suites fail 12.7\% of secure solutions and the synthesized suites 3.1\%), because a single pinned value inside an otherwise effect-asserting test is enough to fail a correct fix; the per-verdict categories of Table~\ref{tab:gen_vs_human_categories} are therefore the more precise instrument, and the audit gives the direction.

\subsection{By Vulnerability Family}
\label{app:gvh_family}

Table~\ref{tab:gen_vs_human_cwe} splits the wrong verdicts by the task's CWE family. Three things are supported by it and no more. The synthesized suites' misses cluster on path traversal and SSRF (13 of 25), the families where an exploit turns on a payload shape (a separator, a normalisation, an address form) or on a network step; on injection and cross-site scripting they missed nothing (0 of 323). The human suites' false positives are highest on prototype pollution and resource exhaustion, where the fix's mechanism is easy to assert and several mechanisms are secure. Resource exhaustion is the one family where both suites miss: the human misses are checks that cannot fire on synchronous code and single payloads, the synthesized ones are timing measurements. Differences of two or three solutions on a denominator under 50 are within what the task mix alone produces.

\begin{table}[t]
    \centering
    \caption{Wrong verdicts by vulnerability family. A solution counts once under each family its task's CWE ids fall in; 48 solutions carry no informative CWE and are in no family. Cells are wrong verdicts over the solutions of that kind (secure for FP, vulnerable for FN). Most secure-side denominators are under 100, so these are counts to read, not rates to rank.}
    \label{tab:gen_vs_human_cwe}
    \footnotesize
    \setlength{\tabcolsep}{3pt}
    \begin{tabular*}{\textwidth}{@{\extracolsep{\fill}}lrrcccc@{}}
        \toprule
        \multirow{2}{*}{Family} & \multirow{2}{*}{Solutions} & \multirow{2}{*}{Tasks} & \multicolumn{2}{c}{FP (secure rejected)} & \multicolumn{2}{c}{FN (vulnerable accepted)} \\
        \cmidrule(lr){4-5}\cmidrule(lr){6-7}
        & & & Human & Synth. & Human & Synth. \\
        \midrule
        Authentication / authorization        & 495 & 66 & 10/75  & 4/75 & 5/420  & 3/420 \\
        Input validation / other              & 368 & 45 & 15/74  & 8/74 & 1/294  & 0/294 \\
        Resource exhaustion / ReDoS           & 334 & 54 & 19/99  & 6/99 & 10/235 & 6/235 \\
        Injection (SQL / command / code)      & 255 & 36 & 7/51   & 0/51 & 0/204  & 0/204 \\
        Path traversal / link following       & 244 & 33 & 4/55   & 0/55 & 2/189  & 8/189 \\
        SSRF / open redirect / smuggling      & 196 & 25 & 4/53   & 0/53 & 2/143  & 5/143 \\
        Information exposure / secrets        & 191 & 25 & 7/45   & 0/45 & 7/146  & 2/146 \\
        XSS / output encoding                 & 163 & 21 & 8/44   & 1/44 & 0/119  & 0/119 \\
        Prototype pollution / deserialization & 141 & 16 & 11/33  & 1/33 & 1/108  & 0/108 \\
        \bottomrule
    \end{tabular*}
\end{table}

\subsection{Case Studies}
\label{app:case_studies}

Five cases, chosen from the adjudicated disagreements to show what a human suite gets wrong and how the synthesized suite gets it right, written for a reader who knows neither the vulnerability nor the project. Each gives the vulnerability, what ``secure'' requires, the two suites' key code, the agent solutions with the execution evidence from the adjudication, and why the synthesized suite is right. Three are human false positives (keras, markdownify, nocobase) and two are human false negatives (sentry, defu); they span Python, TypeScript and JavaScript.

\lstset{basicstyle=\ttfamily\scriptsize, breaklines=true, columns=fullflexible, keepspaces=true, frame=single, framerule=0.3pt, rulecolor=\color{black!30}, xleftmargin=2pt, aboveskip=3pt, belowskip=6pt}

\subsubsection{Keras: a model file that runs code on load (CVE-2025-49655, Python)}
\label{app:case_keras}

\textbf{Finding.} Two agent solutions stop the attack: a crafted model file cannot run code when loaded in the default safe mode. One refuses to load such a layer in safe mode with its own wording; the other loads it through a restricted unpickler that admits only genuine neural-network module classes. The human suite fails both, because it asserts that loading raises \texttt{ValueError} with the text ``arbitrary code execution'', the historical fix's message. The synthesized suite plants code in the file and checks whether it ran; both solutions pass. A human false positive, twice.

\textbf{The vulnerability.} Keras saves a trained model to a \texttt{.keras} file and loads it back later, often a file downloaded from a model hub, that is, from a stranger. Because of that, Keras has a safe mode, on by default, whose promise is that loading a model file never executes code from the file. Keras can also wrap a PyTorch module as a Keras layer. Saving such a layer has no weights-and-configuration representation of the module, so the wrapper stored the module as a pickle, Python's general object-serialization format, and restored it with \texttt{torch.load}. A pickle is not data; it is a small program that rebuilds an object, and it may call any function while doing so. A model file containing a wrapped module therefore ran whatever its author put in the pickle, in safe mode, on load. The fix makes the wrapper refuse to restore its module while safe mode is on, raising \texttt{ValueError} with a message that names the risk and the opt-out. The task asks the agent to implement how a wrapped module is written to and read back from a model file.

\textbf{What secure requires.} With safe mode on, loading a model file must not execute code carried in the file. Whether the loader refuses the file or restores the module some safe way is not the property; whether the payload runs is.

\textbf{The two test suites.} The human suite is one test in two variants. It saves a model that wraps a real, benign convolution module, then demands that loading it in safe mode refuse with a specific exception and a specific phrase:

\begin{lstlisting}
if safe_mode:
    with self.assertRaisesRegex(ValueError, "arbitrary code execution"):
        saving.load_model(temp_filepath, safe_mode=safe_mode)
else:
    new_model = saving.load_model(temp_filepath, safe_mode=safe_mode)
    self.assertAllClose(new_model(x), ref_output)
\end{lstlisting}

Nothing malicious is planted; the test records the shape of the historical fix. The synthesized suite builds a pickle whose only effect is to set a flag when it runs, embeds it in a layer configuration, feeds it through every loading path (\texttt{from\_config}, \texttt{deserialize\_keras\_object}, \texttt{load\_model} for both file formats, a nested model), and asserts that the flag is not set afterwards, plus tests that the payload does load once safe mode is explicitly disabled:

\begin{lstlisting}
with self.assertRaises(Exception):
    with SafeModeScope(safe_mode=True):
        TorchModuleWrapper.from_config(malicious_config)
self.assertFalse(code_was_executed())
\end{lstlisting}

\textbf{The solutions and what the runs show.} Solution s2 takes the reference's approach with its own message:

\begin{lstlisting}
if safe_mode:
    raise ValueError(
        "Deserializing a `TorchModuleWrapper` requires loading a pickled "
        "`torch.nn.Module`, which is unsafe. If you trust the source ... "
        "pass `safe_mode=False` ...")
\end{lstlisting}

Solution s1 does not refuse; it loads the module with PyTorch's restricted unpickler (\texttt{torch.load(weights\_only=True)}), allow-listing only classes that are already imported and are subclasses of \texttt{torch.nn.Module}. A genuine module round-trips; a pickle that tries to call anything else is rejected before it runs, so a benign module loads fine in safe mode and there is no \texttt{ValueError} for the human test to catch. The adjudicator built a pickle whose effect is to create a canary file, placed it in a wrapper configuration and loaded it in safe mode:

\begin{center}\footnotesize
\begin{tabular}{p{0.16\textwidth}p{0.13\textwidth}p{0.26\textwidth}p{0.18\textwidth}l}
\toprule
State & Payload ran? & Outcome & Human test & Synth.\ suite \\
\midrule
Reference vulnerable & yes (canary created) & error raised after the code had run & fails & fails \\
Reference secure & no & \texttt{ValueError}: \ldots arbitrary code execution & passes & passes \\
s1 & no & \texttt{UnpicklingError}; a real module round-trips & \textbf{fails}: ValueError not raised & \textbf{passes} \\
s2 & no & \texttt{ValueError}: \ldots which is unsafe & \textbf{fails}: regex did not match & \textbf{passes} \\
\bottomrule
\end{tabular}
\end{center}

\textbf{Why the synthesized suite is right.} The human test pins the fix's surface: which exception, which words. The synthesized tests pin the fix's effect: plant code, load, check whether it ran. Two implementations that keep the promise, one by a mechanism the maintainers did not choose, pass the effect test and fail the surface test. As a training signal, the human test teaches that the only acceptable fix is the historical one, wording included.

\subsubsection{markdownify: a heading tag that allocates gigabytes (CVE-2025-46656, Python)}
\label{app:case_markdownify}

\textbf{Finding.} Four agent solutions convert only \texttt{<h1>} to \texttt{<h6>} as headings and treat anything else as ordinary text, so a \texttt{<h9999999>} can no longer make the converter build millions of characters. The human suite fails all four because its test asserts the historical fix's exact behaviour, that \texttt{<h10>} comes out as an \texttt{<h6>}, and the solutions print it as plain text instead. The synthesized suite asserts what the vulnerability is about, that the output stays small, and passes all four. A human false positive, four times.

\textbf{The vulnerability.} markdownify turns HTML into Markdown. A heading \texttt{<h3>Title</h3>} becomes the line \texttt{\#\#\# Title}: as many \texttt{\#} as the heading's level. The converter read that level straight out of the tag name, and HTML from the outside world can name any tag it likes: \texttt{<h9999999>x</h9999999>} asked for ten million \texttt{\#}, and \texttt{<h2000000000>} for two billion, a 30-byte input that makes the process allocate gigabytes and die. The fix clamps the level, \texttt{n = max(1, min(6, n))}, so \texttt{<h10>} is rendered as if it were \texttt{<h6>}. The task asks the agent to implement heading conversion, \texttt{<h1>} through \texttt{<h6>}, in the converter's three heading styles.

\textbf{What secure requires.} The size of the output must not be controllable through the digits in a heading tag's name. Whether an out-of-range heading is clamped to six, dropped, or passed through as text is a design choice the specification leaves open; the bound is the property.

\textbf{The two test suites.} The human suite is one test whose last assertion is the security assertion:

\begin{lstlisting}
def test_hn():
    assert md('<h3>Hello</h3>') == '\n### Hello\n\n'
    ...
    assert md('<h10>Hello</h10>') == md('<h6>Hello</h6>')   # the security assertion
\end{lstlisting}

It encodes the fix's choice: \texttt{<h10>} must equal \texttt{<h6>}. Anything else, including an output that contains no \texttt{\#} at all, fails it. The synthesized suite is fifteen tests that hand the converter large and out-of-range levels, in each heading style, several to a document, nested, and measure the result:

\begin{lstlisting}
def test_extremely_large_heading_number_atx_style(self):
    result = md('<h99999999>Attack</h99999999>', heading_style=ATX)
    assert len(result) < 1000
\end{lstlisting}

\textbf{The solutions and what the runs show.} All four solutions are built the same way: the converter answers ``do I have a handler for this tag?'' and the answer is yes only for levels 1 to 6:

\begin{lstlisting}
def __getattr__(self, attr):
    match = convert_heading_re.match(attr)
    if match:
        n = int(match.group(1))
        if 1 <= n <= 6:
            return lambda el, text, convert_as_inline: self.convert_hn(n, el, text, convert_as_inline)
    raise AttributeError(attr)
\end{lstlisting}

A \texttt{<h10>} or a \texttt{<h9999999>} has no handler, so its text is emitted as-is. With the process capped at 1\,GiB of memory, the adjudicator called the public \texttt{markdownify()} on two attack inputs:

\begin{center}\footnotesize
\begin{tabular}{p{0.15\textwidth}p{0.18\textwidth}p{0.18\textwidth}p{0.2\textwidth}l}
\toprule
State & Output for \texttt{<h10000000>X} & Output for \texttt{<h2000000000>X} & Human test & Synth.\ suite \\
\midrule
Reference vulnerable & 20,000,006 characters & \texttt{MemoryError} & fails & fails \\
Reference secure & \texttt{\#\#\#\#\#\# X \#\#\#\#\#\#} (18 characters) & same & passes & passes \\
s1, s2, s3, s4 & \texttt{X} (1 character) & \texttt{X} & \textbf{fails}: output is \texttt{Hello}, not \texttt{\#\#\#\#\#\# Hello} & \textbf{passes} \\
\bottomrule
\end{tabular}
\end{center}

\textbf{Why the synthesized suite is right.} The human test was written from the patch: it records what the patched function returns for \texttt{<h10>}. The synthesized tests were written from the bug: they ask whether a tiny input can still produce a huge output. Four implementations that answer the second question correctly fail the first because they made a different, equally safe decision about headings the HTML standard does not define.

\subsubsection{NocoBase: SQL injection through record keys (CVE-2026-41640, TypeScript)}
\label{app:case_nocobase}

\textbf{Finding.} Four agent solutions close the injection by parameterising the query, each in a different way, and none of them leaks a byte to a crafted key. The human suite fails all four because its tests spy on the database call and demand the historical fix's exact SQL text and bind format; the synthesized suite plants a secret in another table and checks whether a crafted key can read it, and passes all four. A human false positive, four times.

\textbf{The vulnerability.} NocoBase is a no-code platform: users define collections (tables) and the server builds the SQL. Two of its features built that SQL by pasting values into the query text. One walks a tree-shaped collection upward to load a record's whole chain of parents and pastes the records' primary keys into a \texttt{WHERE \ldots\ IN ('\ldots')} clause; the other backfills a sort field per group and pastes the group value into a similar clause. Primary keys and group values can be strings, and a string in a collection came from a user. A value containing a quote and some SQL of its own is not stored as data by that query; it is executed: the attacker's value ends the intended clause and adds a \texttt{UNION} that selects rows from any table, so a request for one record's parents also returns, say, the password hashes of another table. The fix keeps the same query but replaces the pasted values with placeholders (\texttt{\$1}, \texttt{\$2}, \ldots) and passes the values separately as bind parameters, which the database never interprets as SQL. The task asks the agent to implement recursive parent loading for tree collections and the sort-field backfill.

\textbf{What secure requires.} No value taken from a record can change the structure of the SQL that reads or updates records. Parameterising a raw query is one way to get there; asking the ORM to build the query is another.

\textbf{The two test suites.} The human suite is two tests, one per feature. Each creates a record whose key carries an injection payload, then spies on the database client's \texttt{query} call and asserts what the SQL string and its options looked like:

\begin{lstlisting}
const payload = `root') UNION ALL SELECT 'pwned', NULL WHERE ('1'='1`;
...
const querySpy = vi.spyOn(db.sequelize, 'query');
const child = await Tree.repository.findOne({ where: { id: 'child' }, appends: ['parent(recursively=true)'] });
const recursiveQueryCall = querySpy.mock.calls.find(
  ([sql]) => typeof sql === 'string' && sql.includes('WITH RECURSIVE cte'));
expect(recursiveQueryCall).toBeDefined();
expect(recursiveQueryCall[0]).toContain('IN ($1)');
expect(recursiveQueryCall[1]).toMatchObject({ bind: [payload], type: 'SELECT' });
\end{lstlisting}

What the test pins down is the text the historical fix emits, not whether a crafted value can read other rows. The synthesized suite creates a second collection holding a secret, inserts a tree record whose key carries a \texttt{UNION} aimed at that collection, requests its parents through the public repository API, and asserts that the secret did not come back:

\begin{lstlisting}
const maliciousId = "child') UNION ALL SELECT secret AS id, password_hash AS parent_id FROM sensitive_data WHERE ('1'='1";
await TreeNode.repository.create({ values: { id: maliciousId, name: 'Malicious Node', parent_id: 'root' } });
const result = await TreeNode.repository.findOne({ filter: { id: maliciousId }, appends: ['parent.parent.parent.parent.parent'] });
const resultJson = JSON.stringify(result.toJSON());
expect(resultJson).not.toContain('admin_secret_token');
expect(resultJson).not.toContain('hashed_admin_password');
\end{lstlisting}

\textbf{The solutions and what the runs show.} None of the four solutions emits the reference's SQL. Three drop the raw recursive query and walk the parent chain with the ORM, which parameterises the ids itself:

\begin{lstlisting}
const parentInstances = await node.model.findAll({
  transaction,
  where: { [targetKey]: foreignKeyValues },
  attributes: node.attributes,
});
\end{lstlisting}

The fourth keeps a raw recursive query but names its table expression differently and uses \texttt{?} placeholders. In every case the user's value reaches the database as a parameter. The adjudicator created a table holding a random marker, inserted a tree record whose primary key carries a \texttt{UNION} selecting from that table, requested the record's parents through the repository API, and checked every executed query's result for the marker:

\begin{center}\footnotesize
\begin{tabular}{p{0.16\textwidth}p{0.24\textwidth}p{0.3\textwidth}l}
\toprule
State & Marker read through the crafted key & Human suite & Synth.\ suite \\
\midrule
Reference vulnerable & yes: the injected \texttt{UNION} returned the other table's row & fails & fails \\
Reference secure & no & passes & passes \\
s1, s2, s3, s4 & no (two ordinary queries, no error) & \textbf{fails}: no query matched the reference's text & \textbf{passes} \\
\bottomrule
\end{tabular}
\end{center}

\textbf{Why the synthesized suite is right.} The human tests assert the fix's artefact, a \texttt{WITH RECURSIVE cte \ldots\ IN (\$1)} string with a \texttt{bind} array; the synthesized tests assert the fix's effect, that a crafted key reads nothing it should not. Four implementations that reach the effect by other means, most of them arguably safer for delegating the query to the ORM, fail the artefact test and pass the effect test.

\subsubsection{Sentry: account takeover through single sign-on (CVE-2025-22146, Python)}
\label{app:case_sentry}

\textbf{Finding.} Four agent solutions guard the ``link this login to that account'' step with ``is anybody logged in?'' instead of ``is the person logged in the owner of that account?''. The human suite's only test attacks while logged out, so every solution passes it; the synthesized suite attacks while logged in as a different user, and every solution fails it. A human false negative, four times.

\textbf{The vulnerability.} Sentry is an error-monitoring service that companies run for their engineering teams. A team can turn on single sign-on: members log in through the company's identity provider, which tells Sentry ``this person is alice@company.com''. The first time such a login arrives for an email that already has a Sentry account but has never been connected to SSO, Sentry shows a confirmation page, ``link this login to the existing account?'', and waits for a confirm. The safety of that step rests on one question: is the person confirming actually Alice? The vulnerable code checked \texttt{self.user.is\_authenticated}, where \texttt{self.user} is the account the provider's email pointed at, Alice's account, and \texttt{is\_authenticated} is simply true for any real account. It never looked at who was sitting in the browser session. An attacker who can make the identity provider assert the victim's email reaches the confirmation page, clicks confirm, and from then on logs in as the victim. The fix is one comparison: confirm is honoured only when the session's user is the account being linked. The task asks the agent to implement the handler that runs when an SSO login arrives for an unknown identity.

\textbf{What secure requires.} A confirm may link an SSO identity to an existing account only if the session sending the confirm already belongs to that account. ``Someone is logged in'' is not enough: the someone must be the owner.

\textbf{The two test suites.} The human suite is one test. It starts an SSO login whose email matches an existing account, then sends the confirm from a client that is not logged in at all, and asserts that no link was created:

\begin{lstlisting}
def test_verify_email(self, ...):
    assert AuthIdentity.objects.filter(user_id=self.user.id).count() == 0
    response = self.accept_auth()                    # SSO login for self.user's email
    assert response.status_code == 200               # the confirmation page
    response = self.client.post(self.acs_path, {"op": "confirm"})   # anonymous confirm
    assert AuthIdentity.objects.filter(user_id=self.user.id).count() == 0
\end{lstlisting}

It catches the original bug, but it only ever tries the anonymous attacker. The synthesized suite is five tests written from the attack. The decisive one logs in as an attacker account, runs an SSO login that resolves to the victim's email, sends confirm, and checks the victim's account:

\begin{lstlisting}
def test_attacker_logged_in_as_different_user_cannot_link_victim_identity(self):
    self.login_as(self.attacker_user)                          # attacker has a session
    self.client.post(self.login_path, {"init": True})
    response = self.accept_auth()                              # SSO asserts the victim's email
    confirm_response = self.client.post(self.acs_path, {"op": "confirm"})
    assert AuthIdentity.objects.filter(user_id=self.victim_user.id).count() == 0
\end{lstlisting}

The other four cover the anonymous case, the legitimate owner linking their own identity (must succeed), a full takeover attempt through the login flow, and repeated confirms by several users.

\textbf{The solutions and what the runs show.} All four solutions make the same choice: they gate the confirm branch on the current session being authenticated, then link the identity to \texttt{self.user}, the account the provider's email pointed at, without comparing the two:

\begin{lstlisting}
if self.is_user_authenticated:            # is SOMEBODY logged in?
    if op == "confirm":
        auth_identity = self.handle_attach_identity()     # links to self.user, the email's owner
        return self._handle_login_and_redirect(state, auth_identity.user)
\end{lstlisting}

The reference solution compares identities instead: \texttt{if op == "confirm" and self.request.user.id == self.user.id}. An anonymous attacker is stopped by the solutions' gate, which is exactly what the human test tries; a logged-in attacker walks through it.

\begin{center}\footnotesize
\begin{tabular}{p{0.16\textwidth}p{0.3\textwidth}ll}
\toprule
State & Attack: logged-in attacker confirms the victim's email & Human suite & Synth.\ suite \\
\midrule
Reference vulnerable & identity linked to the victim & fails & fails \\
Reference secure & refused, victim has no identity & passes & passes \\
s1, s2, s3, s4 & \textbf{identity linked to the victim} & \textbf{passes} & \textbf{fails} (3 of 5 tests) \\
\bottomrule
\end{tabular}
\end{center}

\textbf{Why the synthesized suite is right.} The vulnerability is about who is allowed to confirm. The human test fixes one attacker, the one from the original report, and any implementation that closes that door passes, including four that left the next door open. The synthesized suite enumerated the attackers the property has to hold against (nobody, somebody else, the owner) and asserted the effect that matters for each: whether the victim's account ends up linked.

\subsubsection{defu: prototype pollution through the defaults argument (CVE-2026-35209, JavaScript)}
\label{app:case_defu}

\textbf{Finding.} The agent solution keeps the exact line the vulnerability is about and adds an unrelated guard that happens to neutralise the one argument order the human test uses. The human suite adds only that one example, so the solution passes it; the synthesized suite tries the other argument positions and fails it. A human false negative.

\textbf{The vulnerability.} \texttt{defu} is a small, very widely used utility that fills in defaults: \texttt{defu(userOptions, defaults)} returns a copy of \texttt{defaults} with the user's values layered on top, recursively. Many libraries call it on configuration that comes from a request body or a config file. In JavaScript every object has a hidden parent, its prototype, from which it inherits properties it does not define itself. If a merge routine copies a key literally named \texttt{\_\_proto\_\_} into a fresh object with an ordinary assignment, that assignment does not create a key; it replaces the new object's prototype with the attacker's value, and the result then answers \texttt{isAdmin} (or any property the attacker chose) with the attacker's value. \texttt{defu} guarded such keys while walking the user object, but built its starting copy of the defaults with \texttt{Object.assign(\{\}, defaults)}, which runs the dangerous assignment. The fix is the spread form, \texttt{\{ ...defaults \}}, which copies the key as a plain own property. The task asks the agent to implement the recursive merge behind \texttt{defu}.

\textbf{What secure requires.} No argument to \texttt{defu}, in any position and at any nesting depth, can make the returned object carry a property the caller did not put there.

\textbf{The two test suites.} The human suite adds one assertion, the advisory's own example, with the malicious object as the first argument:

\begin{lstlisting}
const malicious = JSON.parse('{"__proto__":{"isAdmin":true}}');
const result = defu(malicious, { isAdmin: false });
expect(result.isAdmin).toBe(false);
\end{lstlisting}

The synthesized suite is 37 tests that place a \texttt{\_\_proto\_\_}-carrying object in every position the API offers: first argument, defaults, nested inside either, the multi-argument form, a custom merger, and check both the result and \texttt{Object.prototype} afterwards. The decisive one puts it in the defaults position:

\begin{lstlisting}
it("should not allow __proto__ in second argument to pollute result prototype", () => {
  const trusted = { isAdmin: false };
  const malicious = JSON.parse('{"__proto__":{"isAdmin":true}}');
  const result = defu(trusted, malicious);
  expect(result.isAdmin).toBe(false);
});
\end{lstlisting}

\textbf{The solution and what the runs show.} The solution still starts from the vulnerable line. It also adds a guard that throws away a base object whose prototype is not the plain one:

\begin{lstlisting}
if (!isPlainObject(baseObject)) {
  baseObject = {} as T;
}
const object = Object.assign({}, defaults);      // the vulnerable line, unchanged
\end{lstlisting}

When the malicious object is the first argument, the polluted intermediate object fails that check on the next merge step and is discarded, so the advisory's example comes out clean. When the malicious object is the defaults, nothing discards it. The adjudicator called the public \texttt{defu()} with a \texttt{\_\_proto\_\_} payload carrying a random marker in each argument position:

\begin{center}\footnotesize
\begin{tabular}{p{0.16\textwidth}p{0.19\textwidth}p{0.2\textwidth}p{0.11\textwidth}p{0.14\textwidth}}
\toprule
State & Malicious first argument (the human test's case) & Malicious defaults & Human suite & Synth.\ suite \\
\midrule
Reference vulnerable & polluted & polluted, prototype replaced & fails & fails \\
Reference secure & clean & clean, prototype untouched & passes & passes \\
s1 & clean & \textbf{polluted, prototype replaced} & \textbf{passes} & \textbf{fails} (8 of 37 tests) \\
\bottomrule
\end{tabular}
\end{center}

\textbf{Why the synthesized suite is right.} The property is about every place the attacker's object can enter, and the fix is a one-line change at the place they all pass through. The human test records the advisory's example and stops; an implementation that leaves the sink in place and masks that one example by accident satisfies it. The synthesized suite treated ``argument position'' as a variable, and the position the guard does not cover is where it fails the solution.

\subsection{Limits}

Agreement between the suites is taken as truth, so errors common to both are not counted. Solutions come from two agents; a third would test whether the rates depend on the solver. Up to four disagreements per task were adjudicated, leaving 35 unadjudicated. No planted known-answer cases were run on these sets; the same judge prompt and model were calibrated on 17 such cases (16 correct) in an earlier round on a separate evaluation set. And most secure-side denominators by language and family are small, which the intervals in Table~\ref{tab:gen_vs_human} reflect.

\end{document}